\documentclass[12pt,showpacs,showkeys]{revtex4}

\usepackage{amssymb}
\usepackage{amsmath}
\usepackage{graphicx}
\usepackage{amsbsy}

\newcommand{\bq}{\begin{eqnarray}}
\newcommand{\eq}{\end{eqnarray}}
\newcommand{\bqn}{\begin{eqnarray*}}
\newcommand{\eqn}{\end{eqnarray*}}

\begin{document}
\title{The thermodynamic instabilities of a binary mixture of sticky hard
spheres}

\author{Riccardo Fantoni\footnote{e-mail: {\rm rfantoni@unive.it}}}
\author{Domenico Gazzillo\footnote{e-mail: {\rm gazzillo@unive.it}}}
\author{Achille Giacometti\footnote{e-mail: {\rm achille@unive.it}}}
\affiliation{Istituto Nazionale per la Fisica della Materia and
Dipartimento di Chimica Fisica, Universit\`a di Venezia, S. Marta DD
2137, I-30123 Venezia, Italy}
\date{\today}

\begin{abstract}
\noindent 
The thermodynamic instabilities of a binary mixture of sticky hard
spheres (SHS) in the modified Mean Spherical Approximation (mMSA) and the
Percus-Yevick (PY) approximation are investigated using an approach devised
by X. S. Chen and F. Forstmann \ [\textsl{J. Chem. Phys.} \textbf{97}, 3696
(1992)]. This scheme hinges on a diagonalization of the matrix of second
functional derivatives of the grand canonical potential with respect to the
particle density fluctuations. The zeroes of the smallest eigenvalue and the
direction of the relative eigenvector characterize the instability uniquely.
We explicitly compute three different classes of examples. For a symmetrical
binary mixture, analytical calculations, both for mMSA and for PY, predict
that when the strength of adhesiveness between like particles is smaller
than the one between unlike particles, only a pure condensation spinodal
exists; in the opposite regime, a pure demixing spinodal appears at high
densities. We then compare the mMSA and PY results for a mixture where like
particles interact as hard spheres (HS) and unlike particles as SHS, and for
a mixture of HS in a SHS fluid. In these cases, even though the mMSA and PY
spinodals are quantitatively and qualitatively very different from each
other, we prove that they have the same kind of instabilities. Finally, we
study the mMSA solution for five different mixtures obtained by setting the
stickiness parameters equal to five different functions of the hard sphere
diameters. We find that four of the five mixtures exhibit very different
type of instabilities. Our results are expected to provide a further step
toward a more thoughtful application of SHS models to colloidal fluids.
\end{abstract}

\maketitle

\section{Introduction}

Thermodynamic instabilities are important to locate on the phase diagram of
a fluid system those regions where the system can not exist as a single
phase.

For a one-component system with Helmholtz free energy $A$, pressure $P$, in
a volume $V$, at a temperature $T$, the condition for phase stability is 
$\left( \partial ^{2} A / \partial V^{2} \right) _{T,N}=-\left( 
\partial P /\partial V\right) _{T,N}=1 / \left(V\chi_{T} \right)>0$.
The points where the isothermal compressibility $\chi _{T}$ diverges define
the so called spinodal line, or phase instability boundary \cite{Rowlinson},
that separates the stable from the unstable region of the phase diagram. In
the stable region, where $\chi _{T}>0$, the system can exist in a single
phase, while inside the other region the free energy can be lowered by phase
separation into two phases with different densities. This kind of
instability is usually referred to as mechanical instability, associated
with a gas-liquid transition or condensation
\cite{Rowlinson,Lupis,Gazzillo94,Gazzillo95}. 

In a binary mixture the situation is more complex
\cite{Rowlinson,Lupis,Chen92,Ursenbach94,Gazzillo94,Gazzillo95}. The 
thermodynamic instability is located on the points of the phase diagram
where $\left(\partial^{2}G/\partial x^{2}\right)_{T,P,N}/\chi_T=0$,
where $x$ is the concentration of one of the two species, and $G$ is
the Gibbs free energy. The points where $\chi _{T}^{-1}=0$ are
instabilities of pure condensation (and the Bhatia 
Thornton \cite{Bhatia70} density-density structure factor, $S_{\rho
\rho }(k)$ , diverges at $k=0$). The
points where $(\partial ^{2}G/\partial x^{2})_{T,P,N}=0$ are again
instabilities of pure condensation when 
$\delta =\rho (v_{1}-v_{2})=\left(\partial V/\partial x\right)_{T,P,N}/V$
diverges ($\rho $ is the total number density, $v_{i}$ the partial molar
volume, per particle, of species $i$. In this case all Bhatia Thornton
structure factors diverge at $k=0$) and are instabilities of pure
demixing when $\delta=0$ (in this 
case the Bhatia Thornton concentration-concentration structure factor,
$S_{xx}(k)$ diverges at $k=0$). But, in general (for an asymmetric
mixture), the kind of instability may be in between one of pure condensation
and one of pure demixing, with $\delta $ finite and different from zero
(also in this case all Bhatia Thornton structure factors diverge at
$k=0$). For the particular case of a binary symmetric mixture the only
allowed instabilities are the ones of pure condensation and of pure
demixing, since $\delta =0$.   

A different route was followed by Chen and Forstmann \cite{Chen92} to characterize the
instability uniquely in terms of an angle $\alpha$, function of the density and $x$. 

The purpose of this work is to investigate the nature of instabilities for a
binary mixture of sticky hard spheres (SHS). The SHS one-component model was
originally proposed by Baxter \cite{Baxter68,Baxter71,Watts71}, who showed
how it admitted an analytic solution in the Percus-Yevick (PY)
approximation. The PY solution was later extended to mixtures
\cite{Baxter70,Barboy74,Perram75,Barboy79} and it is nowaday regarded
as extremely useful in colloidal systems. In the SHS model one
accounts for a very short range attractive potential by defining an
infinitely narrow and deep square well. This limit is carried out in a
suitable way so that the second virial coefficient is finite.
Due to its highly
idealized nature, the one-component SHS model is not free of
pathologies \cite{Stell91}. Nonetheless this model has recently regained
considerable attention in studies of colloidal suspensions
\cite{Robertus89,Chen94,Lowen94,Nagele96} especially in its
polydisperse version. 
Since the PY solution of a $p-$component SHS mixture requires the solution
of $p(p+1)/2$ coupled quadratic equations  which are
hard to solve for high $p$, attempts have been made to treat the model with
\textquotedblleft simpler\textquotedblright\ approximations
\cite{Gazzillo00,Gazzillo04}, which would allow analytic solution
even for polydisperse systems. One of these approximations, that we will consider in
this work, is the modified Mean Spherical Approximation (mMSA) \cite{fn1}.

In the present work we apply the Chen and Forstmann formalism to a binary
SHS mixture, using both the mMSA and the PY approximation. The former can be
regarded as the zero density limit of the latter and hence its predictions
must be accepted with care. However it has its main merit in the fact that
it entitles analytical predictions even in the multi-component case, unlike
the PY closure.

Three classes of systems will be discussed in details. First we consider the
symmetric mixture, where equal-size equimolar components interact with
variable strength only in the unlike part. This simplified case was already
studied by Chen and Forstmann for hard core particles with attractive Yukawa
interactions within the reference hypernetted chain approximation. In this
particular SHS case we are able to perform a full characterization of the
mixture both for mMSA and PY. In a second class we discuss two paradigmatic
cases: (i) A fluid having HS interactions among like particles and SHS
interactions for the unlike (System A) and (ii) A fluid formed by one SHS
species and another HS one (System B). For PY both cases have been
previously discussed by Barboy and Tenne \cite{Barboy79}, by Penders
and Vrij \cite{Penders91}, and by Regnaut, Amokrane, and Heno
\cite{Regnaut95,Amokrane97} without, however, 
tackling the issue of the stability nature. Even in these two
cases a detailed analytical investigation can be carried out. Building upon
our recent work \cite{Fantoni05}, we finally discuss a third class of
examples involving a general binary mixture where, however, the stickiness
parameters are related to the sizes of the particles according to some
plausible prescriptions \cite{Fantoni05}. Within the mMSA, we are then able
to discuss the nature of the instabilities previously calculated in Ref. 
\cite{Fantoni05}, by evaluating numerically the Chen and Forstmann
angle $\alpha$. 

The remaining of the paper is organized as follows. In Section \ref{sec:cf}
we briefly outline Chen and Forstmann's approach, in Section
\ref{sec:SHSmixture} we report the PY and mMSA solutions for the
Baxter factor 
correlation function of the SHS mixture. Section \ref{sec:symm} is dedicated
to the binary symmetric mixture, whereas Section \ref{sec:mMSA-PY} to Systems
A and B. Section \ref{subsec:cases} deals with five binary mixtures obtained
setting the stickiness parameters equal to five different functions of the
sphere diameters.

\section{Method for analyzing the instability}

\label{sec:cf} For the sake of completeness, we briefly recall the main
steps of the method reported in Ref. \cite{Chen92}. In doing this, however,
we shall follow the general density functional formalism outlined in Ref. 
\cite{Caillol02} which yields a clearer viewpoint.

\subsection{The Chen and Forstmann formalism}

Consider a binary mixture with $N_{1}$ particles of species 1 with
coordinates ${\mathbf{r}}_{1}^{1},\ldots ,{\mathbf{r}}_{N_{1}}^{1}$
and $N_{2}$ particles of species 2 with coordinates ${\mathbf{r}}_{1}^{2},\ldots ,{\mathbf{r}}_{N_{2}}^{2}$ interacting through spherically symmetric pair
potentials. Define the microscopic densities to be 
\begin{equation}
\pmb{\rho}_{i}({\mathbf{r}})\equiv \sum_{\nu =1}^{N_{i}}\delta ({\mathbf{r}}-
{\mathbf{r}}_{\nu }^{i})~~~i=1,2
\end{equation}
for each one of the two species.

Consider now the non-homogeneous system with an external potential $\phi
_{1}({\mathbf{r}})$ acting on the particles of species 1 and an external
potential $\phi _{2}({\mathbf{r}})$ acting on the particles of species 2.
Let $\mu _{i}$ and $\Lambda _{i}$ be the chemical potential and the de
Broglie thermal wavelength, respectively, for species $i$, $N=N_{1}+N_{2}$
the total number of particles, and ${\mathbf{r}}^{N}=(\{{\mathbf{r}}_{\nu
}^{1}\},\{{\mathbf{r}}_{\nu }^{2}\})$ a short-hand notation for the total
set of coordinates. The grand partition function of the system with total
internal energy $W({\mathbf{r}}^{N})$ is a functional of the generalized
potentials $u_{i}({\mathbf{r}})=\beta \lbrack \mu _{i}-\phi
_{i}({\mathbf{r}})]$
\begin{eqnarray}
\Theta \lbrack u_{1},u_{2}] &\equiv &\sum_{N_{1}=0}^{\infty
}\sum_{N_{2}=0}^{\infty }\frac{1}{\Lambda _{1}^{3N_{1}}N_{1}!}\frac{1}{
\Lambda _{2}^{3N_{2}}N_{2}!}\int e^{-\beta W({\mathbf{r}}^{N})+
\sum_{i=1}^{2}\int u_{i}({\mathbf{r}})\pmb{\rho}_{i}({\mathbf{r}})\;d{
\mathbf{r}}}\;d{\mathbf{r}}^{N}  \notag \\
&=&e^{-\beta \Omega \lbrack u_{1},u_{2}]}~,
\end{eqnarray}
where $\Omega $ is the grand free energy. It can be proven \cite{Caillol02}
that the functional $\Omega $ is strictly concave in $u_{1}$ and $u_{2}$ (if
we opportunely restrict its domain of definition). The equilibrium number
density of species $i$ is given by 
\begin{equation}
\rho _{i}({\mathbf{r}})\equiv \langle \pmb{\rho}_{i}({\mathbf{r}})\rangle =-
\frac{\delta \beta \Omega \lbrack u_{1},u_{2}]}{\delta u_{i}({\mathbf{r}})}~.
\label{intro:rho_nu}
\end{equation}

It follows that the following functional of $\{\rho _{i}\}$ and $\{u_{i}\}$ 
\begin{equation}
\beta A[\rho _{1},\rho _{2},u_{1},u_{2}]\equiv \sum_{i=1}^{2}\int \rho _{i}({
\mathbf{r}})u_{i}({\mathbf{r}})\;d{\mathbf{r}}+\beta \Omega \lbrack
u_{1},u_{2}]~,  \label{intro:A}
\end{equation}
is also strictly concave in $u_{1}$ and $u_{2}$, so it admits a unique
maximum for $u_{i}=\bar{u}_{i}$, $i=1,2$, where the $\{\bar{u}_{i}\}$ can be
determined univocally from Eq. (\ref{intro:rho_nu}) once the equilibrium
densities $\{\rho _{i}\}$ are known.

We now set $\bar{A}[\rho_1,\rho_2]\equiv
A[\rho_1,\rho_2,\bar{u}_1,\bar{u}_2]$. Again one can prove
\cite{Caillol02} that this Helmholtz free energy 
is a strictly convex functional in $\rho_1$ and $\rho_2$.

Introduce the following \textquotedblleft grand free energy
functional\textquotedblright\ of the densities 
\begin{equation}
\beta \Omega ^{\prime }[\rho _{1},\rho _{2}]\equiv \beta \bar{A}[\rho
_{1},\rho _{2}]-\sum_{i=1}^{2}\int \rho _{i}({\mathbf{r}})v_{i}({\mathbf{r}}
)\;d{\mathbf{r}}~,
\end{equation}
where $\{v_{i}\}$ are some given generalized potentials, independent of the
densities. Clearly only when $v_{i}=\bar{u}_{i}$, $i=1,2$, we have $\Omega
^{\prime }=\Omega $, i.e. equilibrium.

Taking the first functional derivative of $\Omega^\prime$ with respect to
the densities we find 
\begin{eqnarray}
\frac{\delta\beta\Omega^\prime[\rho_1,\rho_2]}{\delta\rho_i({\mathbf{r}})}
&=& \frac{\delta\beta\bar{A}[\rho_1,\rho_2]}{\delta\rho_i({\mathbf{r}})}-v_i(
{\mathbf{r}})  \notag \\
&=&\bar{u}_i({\mathbf{r}})-v_i({\mathbf{r}})~,
\end{eqnarray}
where in the second equality Eqs. (\ref{intro:rho_nu}) and (\ref{intro:A})
where used. At equilibrium we then have that the first functional
derivatives of $\Omega^\prime$ vanish and $\Omega^\prime$ attains its
minimum value.

The second functional derivatives of $\Omega ^{\prime }$ with respect to the
densities at equilibrium are \cite{Caillol02} 
\begin{equation}
\left. \frac{\delta ^{2}\beta \Omega ^{\prime }[\rho _{1},\rho _{2}]}{\delta
\rho _{i}({\mathbf{r}}_{1})\delta \rho _{j}({\mathbf{r}}_{2})}\right\vert
_{equil.}=\left. \frac{\delta \bar{u}_{i}({\mathbf{r}}_{1})}{\delta \rho
_{j}({\mathbf{r}}_{2})}\right\vert _{equil.}=\frac{\delta _{ij}\delta ({
\mathbf{r}}_{1}-{\mathbf{r}}_{2})}{\rho _{i}({\mathbf{r}}_{1})}-c_{ij}({
\mathbf{r}}_{1},{\mathbf{r}}_{2})~,
\end{equation}
where $c_{ij}({\mathbf{r}}_{1},{\mathbf{r}}_{2})$ are the partial direct
correlation functions of the system.

So a Taylor expansion, up to the second order terms, yields the fluctuation
of $\Omega ^{\prime }$ around the equilibrium caused by small density
fluctuations 
\begin{eqnarray}
\delta \Omega ^{\prime } &=&\Omega ^{\prime }[\rho _{1}+\delta \rho
_{1},\rho _{2}+\delta \rho _{2}]-\Omega ^{\prime }[\rho _{1},\rho _{2}] 
\notag  \label{intro:tayloromegap} \\
&=&\frac{1}{2\beta }\int \int \sum_{i,j}\left[ \frac{\delta _{ij}\delta ({
\mathbf{r}}_{1}-{\mathbf{r}}_{2})}{\rho _{i}({\mathbf{r}}_{1})}-c_{ij}({
\mathbf{r}}_{1},{\mathbf{r}}_{2})\right] \delta \rho _{i}({\mathbf{r}}
_{1})\delta \rho _{j}({\mathbf{r}}_{2})\;d{\mathbf{r}}_{1}d{\mathbf{r}}_{2}~.
\end{eqnarray}

If the system is homogeneous and isotropic at equilibrium
(i.e. $\bar{u}_{i}({\mathbf{r}})=\beta \mu _{i}$, $i=1,2$), so that  
\begin{eqnarray}
\rho _{i}({\mathbf{r}}) &=&\frac{N_{i}}{V}=\rho _{i}~, \\
c_{ij}({\mathbf{r}}_{1},{\mathbf{r}}_{2}) &=&c_{ij}(|{\mathbf{r}}_{1}-{
\mathbf{r}}_{2}|)~,
\end{eqnarray}
where $V$ is the volume (assumed large enough), then we can rewrite the
integral of \ Eq. (\ref{intro:tayloromegap}), which is a convolution,
as a ${\mathbf{k}}$-integral of a product of Fourier
transforms. Replacing the ${\mathbf{k}}$-integral $\left[ \left( 2\pi
\right) ^{-3}\int d{\mathbf{k}}\ldots \right] $ by a sum over discrete
${\mathbf{k}}$-values $\left[V^{-1}\sum_{{\mathbf{k}}}\ldots \right]$,
one obtains  
\begin{equation}
\delta \Omega ^{\prime }=\frac{1}{2\beta }\frac{1}{V}\sum_{{\mathbf{k}}
}\sum_{i,j}\delta \bar{\rho}_{i}^{\star }({\mathbf{k}})\tilde{A}
_{ij}(k)\delta \bar{\rho}_{j}({\mathbf{k}})~,  \label{intro:op1}
\end{equation}
where $\delta \bar{\rho}_{i}({\mathbf{k}})=\delta
\tilde{\rho}_{i}({\mathbf{k}})/\sqrt{\rho _{i}}$ and the asterisk
indicates complex conjugation, having 
denoted with the tilde the Fourier transform 
\begin{equation}
\tilde{f}({\mathbf{k}})\equiv
\int_{V}f({\mathbf{r}})e^{i{\mathbf{k}}\cdot {\mathbf{r}}}\;d{\mathbf{r}}~,
\end{equation}
so that 
\begin{equation}
\tilde{A}_{ij}(k)=\delta _{ij}-\sqrt{\rho _{i}\rho _{j}}\;\tilde{c}_{ij}(k)~.
\label{intro:tildeA}
\end{equation}
Notice that, due to the symmetry of the direct correlation functions under
exchange of species indexes, the matrix $\mathbf{\tilde{A}}(k)$ is symmetric.

The probability distribution for the density fluctuations $\delta \rho _{i}$
(at constant $T$, $V$, and $\{\mu _{i}\}$) is proportional to $e^{-\beta
\delta \Omega ^{\prime }}$ \cite{Landau}. We therefore get for the mean
values of the fluctuation products 
\begin{eqnarray}
\langle \delta \bar{\rho}_{i}^{\star }({\mathbf{k}})\delta \bar{\rho}_{j}({
\mathbf{k}})\rangle &=&V\left[ \mathbf{\tilde{A}}^{-1}\right] _{ij}(k) 
\notag \\
&=&V[\delta _{ij}+\sqrt{\rho _{i}\rho _{j}}\;\tilde{h}_{ij}(k)]~,
\end{eqnarray}
where the last equality exploits the Ornstein-Zernike (OZ) equations between
the partial total correlation functions $h_{ij}$ and the partial direct
correlation functions.

Next define the molar fraction of species $i$ to be $x_{i}=\rho _{i}/\rho
, $ with $\rho =\sum_{i}\rho _{i}$ being the total density of the mixture.
One usually introduces \cite{Bhatia70} two linear combinations of
fluctuations of partial densities, i.e. the fluctuation of total
density, $\delta \tilde{\rho}({\mathbf{k}})$, and the fluctuation of
concentration of species 1, $\delta \tilde{x}({\mathbf{k}})$, 
\begin{eqnarray} \label{intro:tilderho}
\delta \tilde{\rho}({\mathbf{k}}) &=&\delta \tilde{\rho}_{1}({\mathbf{k}}
)+\delta \tilde{\rho}_{2}({\mathbf{k}})  \\
&=&\sqrt{\rho }[\sqrt{x_{1}}\delta \bar{\rho}_{1}({\mathbf{k}})+\sqrt{x_{2}}
\delta \bar{\rho}_{2}({\mathbf{k}})]~,  \notag \\ \label{intro:tildex} 
\delta \tilde{x}({\mathbf{k}}) &=&\frac{1}{\rho ^{2}}[\rho _{2}\delta \tilde{
\rho}_{1}({\mathbf{k}})-\rho _{1}\delta \tilde{\rho}_{2}({\mathbf{k}})] \\
&=&\sqrt{\frac{x_{1}x_{2}}{\rho }}[\sqrt{x_{2}}\delta \bar{\rho}_{1}({
\mathbf{k}})-\sqrt{x_{1}}\delta \bar{\rho}_{2}({\mathbf{k}})]~,  \notag
\end{eqnarray}
so that, if $\delta \tilde{\rho}_{1}$ and $\delta \tilde{\rho}_{2}$ change
in proportion to their respective mean concentration, then $\delta \tilde{x}
=0$.

We also introduce 
\begin{eqnarray}
\delta \bar{\rho}({\mathbf{k}}) &=&\frac{1}{\sqrt{\rho }}\delta \tilde{\rho}(
{\mathbf{k}})~, \\
\delta \bar{x}({\mathbf{k}}) &=&\sqrt{\frac{\rho }{x_{1}x_{2}}}\delta \tilde{
x}({\mathbf{k}})~,
\end{eqnarray}
so that, in terms of the following two column vectors 
\begin{equation}
{\mathbf{u}}({\mathbf{k}})=\left( 
\begin{array}{c}
\delta \bar{\rho}_{1}({\mathbf{k}}) \\ 
\delta \bar{\rho}_{2}({\mathbf{k}})
\end{array}
\right) ~,~~~{\mathbf{v}}({\mathbf{k}})=\left( 
\begin{array}{c}
\delta \bar{\rho}({\mathbf{k}}) \\ 
\delta \bar{x}({\mathbf{k}})
\end{array}
\right) ~,
\end{equation}
Eqs. (\ref{intro:tilderho}) and (\ref{intro:tildex}) can be written in
compact notation as ${\mathbf{u}}=\mathbf{U}{\mathbf{v}}$ where 
\begin{equation}
\mathbf{U}=\left( 
\begin{array}{cc}
\sqrt{x_{1}} & \sqrt{x_{2}} \\ 
\sqrt{x_{2}} & -\sqrt{x_{1}}
\end{array}
\right) ~,
\end{equation}
notice that $\mathbf{U}^{2}=\mathbf{I}$, where $\mathbf{I}$ is the identity
matrix.

We find then from Eq. (\ref{intro:op1}) (superscript $T$ indicating the
transpose) 
\begin{equation}
\delta \Omega ^{\prime }=\frac{1}{2\beta }\frac{1}{V}\sum_{{\mathbf{k}}}{
\mathbf{v}}^{T\star }({\mathbf{k}})\mathbf{M}(k){\mathbf{v}}({\mathbf{k}})~,
\label{intro:op2}
\end{equation}
where $\mathbf{M}(k)$ is the following symmetric matrix 
\begin{equation}
\mathbf{M}(k)=\mathbf{U}\mathbf{\tilde{A}}(k)\mathbf{U}=\left( 
\begin{array}{cc}
M_{\rho \rho } & M_{\rho x} \\ 
M_{x\rho } & M_{xx}
\end{array}
\right) ~,
\end{equation}
with 
\begin{eqnarray} \label{intro:Mc1}
M_{\rho \rho } &=&1-\rho \lbrack x_{1}^{2}\tilde{c}_{11}+x_{2}^{2}\tilde{c}
_{22}+2x_{1}x_{2}\tilde{c}_{12}]~,  \\ \label{intro:Mc2}
M_{xx} &=&1-\rho x_{1}x_{2}[\tilde{c}_{11}+\tilde{c}_{22}-2\tilde{c}_{12}]~,
\\  \label{intro:Mc3}
M_{\rho x} &=&M_{x\rho }=\rho \sqrt{x_{1}x_{2}}[x_{2}\tilde{c}_{22}-x_{1}
\tilde{c}_{11}-(x_{2}-x_{1})\tilde{c}_{12}]~.
\end{eqnarray}
The elements of the $\mathbf{M}(0)$ matrix are related to thermodynamic
quantities \cite{Chen92}, as shown in the Appendix. In particular the
determinant of $\mathbf{M}$ is 
\begin{equation}
\det (\mathbf{M})=x_{1}x_{2}\frac{(\chi _{T}^{0})^{2}}{\chi _{T}V}\left( 
\frac{\partial ^{2}G}{\partial x_{1}^{2}}\right) _{T,P,N}~,
\label{det-thermo}
\end{equation}
where $\chi _{T}$ is the isothermal compressibility, and $\chi _{T}^{0}=\beta
/\rho $ is the isothermal compressibility of the ideal gas.

For the particular systems that we shall consider in the following, it turns
out that the matrix $\mathbf{\tilde{A}}$ can be written, using the
Wiener-Hopf factorization in terms of the Baxter factor matrix
$\widehat{\mathbf{Q}}$ \cite{Baxter70}  
\begin{equation}
\mathbf{\tilde{A}}(k)=\widehat{\mathbf{Q}}^{T\star }(k)\widehat{\mathbf{Q}}
(k)~.  \label{intro:wh}
\end{equation}
Hence $\det [\mathbf{M}(k)]=\det [\mathbf{\tilde{A}
}(k)]=|\det [\widehat{\mathbf{Q}}(k)]|^{2}\geq 0$ and $\mathrm{trace}[
\mathbf{M}(k)]=\mathrm{trace}[\mathbf{\tilde{A}}(k)]\geq 0$.

The inverse of $\mathbf{M}(k)$ yields the mean square fluctuations of total
density and concentration, i.e. the density-density structure factor
$S_{\rho\rho}(k)$, the concentration-concentration structure factor
$S_{xx}(k)$, and the cross term $S_{\rho x}(k)$ \cite{Bhatia70}  
\begin{eqnarray}  \label{intro:MS1}
S_{\rho\rho}(k)&=&\frac{1}{V} \langle\delta\bar{\rho}^\star({\mathbf{k}}
)\delta\bar{\rho}({\mathbf{k}})\rangle
=[\mathbf{M}^{-1}]_{\rho\rho}(k)~, \\
\label{intro:MS2}
S_{xx}(k)&=&\frac{x_1x_2}{V} \langle\delta\bar{x}^\star({\mathbf{k}})\delta
\bar{x}({\mathbf{k}})\rangle =x_1x_2\;[\mathbf{M}^{-1}]_{xx}(k)~, \\
\label{intro:MS3}
S_{\rho x}(k)&=&\frac{\sqrt{x_1x_2}}{V} \langle\delta\bar{\rho}^\star({
\mathbf{k}})\delta\bar{x}({\mathbf{k}})\rangle =\sqrt{x_1x_2}\;[\mathbf{M}
^{-1}]_{\rho x}(k)~.
\end{eqnarray}

Now, since $\mathbf{M}(k)$ is a symmetric matrix, it can be diagonalized
through an orthogonal change of basis and it will have real eigenvalues 
\begin{equation}
\lambda _{\pm }(k)=\frac{\mathrm{trace}[\mathbf{M}(k)]\pm \sqrt{\{\mathrm{
trace}[\mathbf{M}(k)]\}^{2}-4\det [\mathbf{M}(k)]}}{2}~,
\end{equation}
with $\lambda _{+}(k)\geq \lambda _{-}(k)\geq 0$. For the normalized
eigenvectors we find 
\begin{equation}
\mathbf{z}_{\pm }(k)=\left( 
\begin{array}{c}
a_{\pm }(k) \\ 
b_{\pm }(k)
\end{array}
\right) ~,  \label{zpm}
\end{equation}
with 
\begin{eqnarray}
a_{\pm }(k) &=&1/\sqrt{1+\left( \frac{M_{\rho \rho }(k)-\lambda _{\pm }(k)}{
M_{\rho x}(k)}\right) ^{2}}~, \\
b_{\pm }(k) &=&-a_{\pm }\frac{M_{\rho \rho }(k)-\lambda _{\pm }(k)}{M_{\rho
x}(k)}~.
\end{eqnarray}

The transition matrix to the base formed by the eigenvectors will be 
\begin{equation}
\mathbf{Z}(k)=\left( 
\begin{array}{cc}
a_{+}(k) & a_{-}(k) \\ 
b_{+}(k) & b_{-}(k)
\end{array}
\right) ~,
\end{equation}
Eq. (\ref{intro:op2}) can then be recast into the form 
\begin{equation}
\delta \Omega ^{\prime }\left( \delta \rho _{1},\delta \rho _{2}\right) =
\frac{1}{2\beta }\frac{1}{V}\sum_{{\mathbf{k}}}\ [\lambda _{+}(k)\ |\delta 
\bar{\rho}_{+}({\mathbf{k}})|^{2}+\lambda _{-}(k)\ |\delta \bar{\rho}_{-}({
\mathbf{k}})|^{2}]~,  \label{intro:op3}
\end{equation}
where $\delta \bar{\rho}_{\pm }$ are the Fourier components of the vector
for the total density and concentration fluctuation in the eigenvector base,
namely 
\begin{equation}
\mathbf{Z}^{-1}{\mathbf{v}}=\left( 
\begin{array}{c}
\delta \bar{\rho}_{+} \\ 
\delta \bar{\rho}_{-}
\end{array}
\right) ~,
\end{equation}
or 
\begin{eqnarray}
\delta \bar{\rho}_{+}\left( {\mathbf{k}}\right) &=&a_{+}(k)\ \delta \bar{\rho
}\left( {\mathbf{k}}\right) +b_{+}(k)\ \delta \bar{x}\left( {\mathbf{k}}
\right) ~, \\
\delta \bar{\rho}_{-}\left( {\mathbf{k}}\right) &=&a_{-}(k)\ \delta \bar{\rho
}\left( {\mathbf{k}}\right) +b_{-}(k)\ \delta \bar{x}\left( {\mathbf{k}}
\right) ~.  \label{e2}
\end{eqnarray}

\subsection{Characterization of the instability}

We wish to know which combination of density and concentration fluctuations, 
$\left( \delta \bar{\rho},\delta \bar{x}\right) $ or $\left( \delta \bar{\rho
}_{+},\delta \bar{\rho}_{-}\right) $, yields the smallest increase $\delta
\Omega ^{\prime }$ of grand free energy. The border of a stability region
(spinodal line) will be determined by the smaller eigenvalue $\lambda
_{-}(k) $ going to zero. It is important to remark that the minimum
eigenvalue will vanish if and only if $\det [\mathbf{M}(k)]=\lambda
_{-}(k)\lambda _{+}(k)=|\det [\widehat{\mathbf{Q}}(k)]|^{2}$ vanishes. The
spinodal equation thus corresponds to

\begin{equation}
\lambda _{-}(k)=0\qquad \text{or\qquad }\det [\widehat{\mathbf{Q}}(k)]=0.
\end{equation}

\begin{figure}[h!]
\begin{center}
\includegraphics[width=10cm]{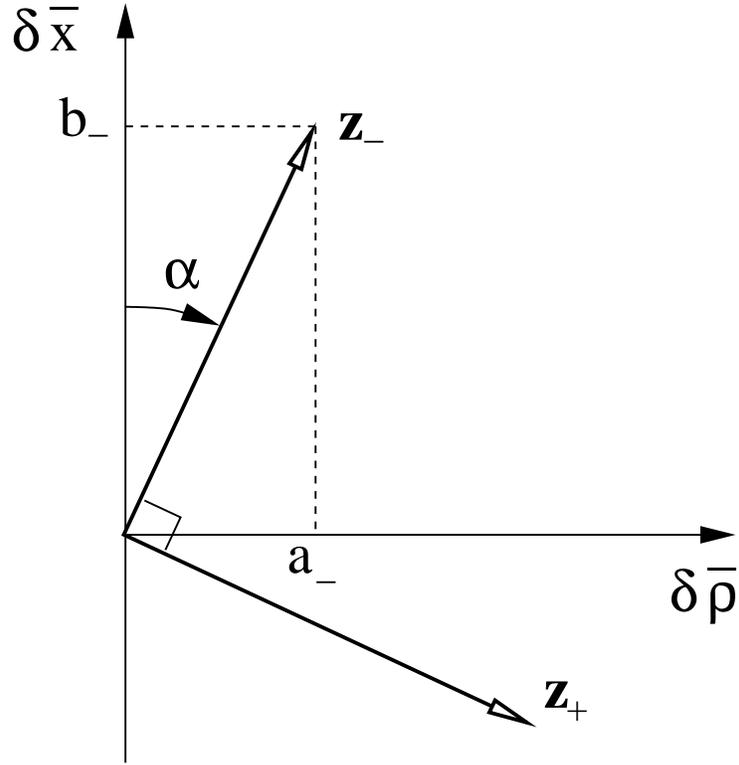}
\end{center}
\caption{Schematic representation of the two orthonormal vectors
${\bf z}_\pm$ defined in Eq. (\ref{zpm}) and of the angle 
$\alpha$ defined in Eq. (\ref{intro:angle}) when
$\alpha\in[0,\pi/2]$. When $\alpha\in[-\pi/2,0]$ the angle
shown in the Fig. corresponds to $|\alpha|$ and $\delta\bar{x}$ to
$-\delta\bar{x}$.}
\label{fig:angle}
\end{figure}

For all $\overline{{\mathbf{k}}}$-vectors with $\bar{k}=\left\vert 
\overline{{\mathbf{k}}}\right\vert $ being a solution of the spinodal
equation, we can calculate the related eigenvector $\mathbf{z}_{-}(\bar{k})$
and find, from Eq. (\ref{e2}), one non-zero linear combination $\delta \bar{
\rho}_{-}(\overline{{\mathbf{k}}})$ of density and concentration
fluctuations for which $\delta \Omega ^{\prime }=0$. Thus $\mathbf{z}_{-}(
\bar{k})=\left[ a_{-}(\overline{k}),b_{-}(\overline{k})\right]^T$
characterizes the phase transition uniquely. On defining the angle (see
Fig. \ref{fig:angle}) 
\begin{equation} \label{intro:angle}
\alpha =\arctan \left( \frac{a_{-}}{b_{-}}\right) _{k=\overline{k}}=\arctan 
\left[ \frac{\widehat{Q}_{12}(\overline{k})\sqrt{x_{1}}-\widehat{Q}_{11}(
\overline{k})\sqrt{x_{2}}}{\widehat{Q}_{12}(\overline{k})\sqrt{x_{2}}+
\widehat{Q}_{11}(\overline{k})\sqrt{x_{1}}}\right] ~,
\end{equation}
the instability will be predominantly of the demixing type when $\alpha $ is
close to $0$ (i.e. only concentration fluctuations occur) and predominantly
of the condensation type when $\alpha $ is close to $\pm \pi /2$ (i.e. only
density fluctuates at fixed concentration). 

The same feature can be seen in real space. When $\lambda _{-}(\bar{k})=0$
and $\alpha =0$ ($\Rightarrow a_{-}=0,\text{ }b_{+}=0$, and
therefore $\delta \bar{\rho}_{+}=a_{+}\ \delta \bar{\rho},\text{ }\delta 
\bar{\rho}_{-}=b_{-}\ \delta \bar{x}$), one can get $\delta \Omega
^{\prime }=0$ only if $\delta \bar{\rho}_{+}(\overline{{\mathbf{k}}})=0$,
which requires $\delta \widetilde{\rho }(\overline{{\mathbf{k}}})=0$, i.e.
the fluctuations that do not increase the \textquotedblleft grand free
energy\textquotedblright\ can be expressed as 
\begin{eqnarray}
\delta \rho _{1}({\mathbf{r}}) &=&\frac{1}{V}\sum_{\overset{\scriptstyle{
\mathbf{k}}}{|{\mathbf{k}}|=\bar{k}}}\delta \rho _{1}({\mathbf{k}})\;e^{-i{
\mathbf{k}}\cdot {\mathbf{r}}}~,  \label{demixing1} \\
\delta \rho _{2}({\mathbf{r}}) &=&-\delta \rho _{1}({\mathbf{r}})~.
\label{demixing2}
\end{eqnarray}
On the other hand, when $\lambda _{-}(\bar{k})=0$ and $\alpha =\pm \pi /2$ $
\left( \Rightarrow \text{ }\delta \bar{\rho}_{+}=b_{+}\ \delta \bar{x},\text{
}\delta \bar{\rho}_{-}=a_{-}\ \delta \bar{\rho}\right) $, $\delta \bar{\rho}
_{+}(\overline{{\mathbf{k}}})=0$ now requires $\delta \widetilde{x}(
\overline{{\mathbf{k}}})=0$, which corresponds to 
\begin{equation}
\delta \rho _{2}({\mathbf{r}})=+\frac{\rho _{2}}{\rho _{1}}\delta \rho _{1}({
\mathbf{r}})~.  \label{condensation2}
\end{equation}
Eq. (\ref{demixing1}) yields oscillating partial density fluctuations for
species 1 on the spinodal line, whereas Eqs. (\ref{demixing2}) and
(\ref{condensation2}) represent the two different behaviors of the
species 2 in 
correspondence to the two extreme values of $\alpha $ (0 and $\pm \pi /2,$
respectively). For $\alpha =0$, the fluctuations of species 2 must be in
opposition of phase compared to those of species 1, [see Eq. (\ref{demixing2})],
and this can be clearly interpreted as related to spatial demixing. In the
opposite case ($\alpha =\pm \pi /2$), Eq. (\ref{condensation2}) means that
an increase of species 1 in some region drives an increase of species 2 in
the same region, a clear indication of a condensation type of instability.
When $\alpha $ varies from zero to $\pm \pi /2$ the allowed fluctuations
will continuously vary from the pure demixing to the pure condensation type.

For a class of approximations (closures) having the partial direct
correlation functions vanishing beyond a finite range, it was proven
in \cite{Baxter70} that $\widehat{\mathbf{Q}}(k)$ is non-singular for
any $k>0$, so 
we can limit our search for the zeroes of the minimum eigenvalue to the case 
$k=0$.  Moreover, since $\lim_{k\rightarrow \infty
}\det [\widehat{\mathbf{Q}}(k)]=1$ and $\det [\widehat{\mathbf{Q}}(k)]$ is a
continuous function of $k$, we must also have $\det
[\widehat{\mathbf{Q}}(k=0)]$ non-negative, otherwise $\det
[\widehat{\mathbf{Q}}(k)]$ would 
vanish for some finite $k$. We can use this last condition to determine
which regions of the phase diagram are unstable. We cannot in fact gather
this information by just looking at the matrix $\tilde{\mathbf{A}}$, which
is always positive definite when non-singular. 
In the following, whenever we omit the dependence from the wave-vector $k$, we 
shall refer to the case $k=0$.

\section{The binary sticky hard sphere fluid}

\label{sec:SHSmixture}

We consider the SHS mixture described in the introduction by the 
following square-well interaction potential between a sphere of
species $i$ and one of species $j$ 
\cite{Baxter68,Baxter71,Perram75,Barboy79}
\begin{equation}
\beta \phi _{ij}(r)=\left\{ 
\begin{array}{ll}
+\infty & 0<r<\sigma _{ij}~, \\ 
\displaystyle-\ln \left( \frac{1}{12\tau _{ij}}\frac{R_{ij}}{R_{ij}-\sigma
_{ij}}\right) & \sigma _{ij}\leq r\leq R_{ij}~, \\ 
0 & r>R_{ij}~,
\end{array}
\right.  \label{SW}
\end{equation}
where $\beta =1/(k_{B}T)$ ($k_{B}$ being Boltzmann constant and $T$ the
temperature), $\sigma _{ij}=(\sigma _{i}+\sigma _{j})/2$ ($\sigma _{i}$
being the diameter of a sphere of species $i$), $R_{ij}-\sigma _{ij}$
denotes the well width, and the dimensionless parameter 
\begin{equation}
\frac{1}{\tau _{ij}}=\frac{\epsilon _{ij}}{\tau }=\frac{\nu _{ij}}{\tau
^{\ast }}\geq 0~,  \label{tauij}
\end{equation}
measures the strength of surface adhesiveness or \textquotedblleft
stickiness\textquotedblright\ between particles of species $i$ and $j$. In
(\ref{tauij}), $\tau $ is an unspecified increasing function of $T$, and we
introduced the dimensionless quantities $\nu _{ij}=\epsilon _{ij}/\epsilon
_{11}$ and $\tau ^{\ast }=\tau /\epsilon _{11}$. The next step which defines
the SHS model consists in taking the \textquotedblleft
sticky\textquotedblright\ limit $\left\{ R_{ij}\right\} \rightarrow \left\{
\sigma _{ij}\right\} $. Notice that the logarithm in the initial square-well
potential (\ref{SW}) is chosen so to have a simple expression for the
Boltzmann factor, which reduces to a combination of an Heaviside step
function and a Dirac delta function in the sticky limit.

Within a class 
of mixed closures for which the partial direct correlation functions
$c_{ij}(r)$ after the sticky limit vanish beyond $\sigma _{ij}$
[generalized 
PY (GPY) approximation \cite{Gazzillo04}], the model can be analytically
solved for the Baxter factor correlation function 
\begin{equation}
q_{ij}(r)=\left\{ 
\begin{array}{ll}
\frac{1}{2}a_{i}(r-\sigma _{ij})^{2}+(b_{i}+a_{i}\sigma _{ij})(r-\sigma
_{ij})+K_{ij}~, & L_{ij}=(\sigma _{i}-\sigma _{j})/2\leq r\leq \sigma _{ij}~,
\\ 
0~, & \text{elsewhere}~,
\end{array}
\right.  \label{m3}
\end{equation}

\begin{equation}
a_{i}=\frac{1}{\Delta }+\frac{3\xi _{2}\sigma _{i}}{\Delta ^{2}}- \frac{
12\zeta _{i}}{\Delta }~,\qquad b_{i}=\left( \frac{1}{\Delta }-a_{i}\right)
\frac{\sigma _{i}}{2}~,  \label{m4}
\end{equation}

\begin{equation}
\xi _{n}=\frac{\pi }{6}\sum_i\rho _{i}\sigma _{i}^{n}~,\qquad \zeta_{i}=
\frac{\pi }{6}\sum_m\rho _{m}\sigma _{m}K_{im}~,\qquad \Delta=1-\xi _{3}~,
\label{m5}
\end{equation}

The Baxter factor matrix $\widehat{\mathbf{Q}}(k)$ first introduced
in Eq. (\ref{intro:wh}) is related to Baxter factor correlation
function through
\bq
\widehat{Q}_{ij}(k)=\delta _{ij}-2\pi \sqrt{\rho _{i}\rho _{j}}\;\widehat{q}
_{ij}(k)~,
\eq
where $\widehat{q}_{ij}(k)$ is the one-dimensional Fourier transform of
$q_{ij}(r)$. It can be expressed in terms of spherical Bessel
functions of the zeroth and first order and its explicit expression
can be found in Eq. (27) of Ref. \cite{Gazzillo00}, and will not be reproduced here.    

The symmetric matrix $K_{ij}$ is given by 
\begin{equation}
K_{ij}=\frac{\sigma _{ij}^{2}}{12\tau _{ij}}\;\bar{y}_{ij}~,  \label{K_ij}
\end{equation}
where $\bar{y}_{ij}=y_{ij}(\sigma _{ij}^{+})$ are the contact values of the
partial cavity functions. For this kind of system a more natural parameter
to use in place of the total density $\rho =\sum_{i}\rho _{i}$ is the total
packing fraction $\eta =\xi _{3}$.

In the modified Mean Spherical Approximation [$c_{ij}(r)=f_{ij}(r)$
when $r>\sigma _{ij}$, where $f_{ij}(r)=\exp \left[ -\beta \phi
_{ij}(r)\right] -1$ 
are the Mayer functions] one can show \cite{Gazzillo04} that \cite{fn2}
\begin{equation}
\bar{y}_{ij}=1~~~\mbox{for all $i$ and $j$}~,
\end{equation}

In the Percus-Yevick approximation [$c_{ij}(r)=f_{ij}(r)y_{ij}(r)$] one can
show that the $\bar{y}_{ij}$ have to satisfy the following set of coupled
quadratic equations \cite{Perram75} 
\begin{equation}
\bar{y}_{ij}\sigma _{ij}=a_{i}\sigma _{ij}+b_{i}+2\pi \sum_{k}\rho _{k}\frac{
\sigma _{kj}^{2}}{12\tau _{kj}}\;\bar{y}_{kj}\;q_{ki}(L_{ki})  \label{PYyij}
\end{equation}

It is worth stressing that the above expressions are valid for both
the mMSA and the PY closures, provided that the correct values of
$\bar{y}_{ij}$ are inserted into the matrix $K_{ij}$ given in
Eq. (\ref{K_ij}) \cite{Gazzillo00,Fantoni05}. All the results gathered
so far in this Section are 
valid for a generic $p-$component SHS mixture. In the rest of the work we
will specialize to two-component ($p=2$) mixtures. For a binary mixture the
determinant of $\widehat{\mathbf{Q}}(0)$ can be reduced to the following
simple expression \cite{fn3}
\begin{eqnarray}
\det [\widehat{\mathbf{Q}}(0)] &=&\frac{1+2\eta }{(1-\eta )^{2}}-\frac{\eta
_{1}\lambda _{11}^{BT}+\eta _{2}\lambda _{22}^{BT}}{(1-\eta )^{2}}-  \notag
\label{detQSHS} \\
&&\frac{\eta _{1}\eta _{2}}{(1-\eta )^{3}}[3(\lambda _{11}^{BT}+\lambda
_{22}^{BT}-2\lambda _{12}^{BT})-\lambda _{11}^{BT}\lambda
_{22}^{BT}+(\lambda _{12}^{BT})^{2}]~,
\end{eqnarray}
where 
\begin{eqnarray}
\eta _{i} &=&\frac{\pi }{6}\rho _{i}\sigma _{i}^{3}~, \\
\lambda _{ij} &=&\frac{\bar{y}_{ij}}{\tau _{ij}}~, \\
\lambda _{ij}^{BT} &=&(1-\eta )\ \lambda _{ij}~\frac{\sigma _{ij}^{2}}{
\sigma _{i}\sigma _{j}}.
\end{eqnarray}

Our task is the determination of the spinodal line and of the nature of the
instability. These can be expressed respectively by the reduced
temperature $\tau^*=f_\tau(\rho\sigma_1^3,x_1,\zeta,\{\nu_{ij}\})$ and
the angle $\alpha=f_\alpha(\rho\sigma_1^3,x_1,\zeta,\{\nu_{ij}\})$,
where $\zeta=\sigma_2/\sigma_1$. Sometimes it also proves convenient
to use another set of independent variables, namely
$\eta,x_1,\zeta,\{\nu_{ij}\}$. 

We anticipate that, while $f_{\tau }$ will in general depend on the
particular chosen closure, $f_{\alpha }$ need not mirror this feature. An
example is the case studied in Section \ref{sec:mMSA-PY}, where two
$\epsilon _{ij}$ are zero and
$\epsilon_{\bar{\imath}\bar{\jmath}}>0$. Then $\lambda _{ij}=0$ for
$i\neq\bar{\imath}$ or $j\neq \bar{\jmath}$ and the spinodal equation 
\begin{equation}
\lambda _{-}(0)=0\qquad \text{or\qquad }\det [\widehat{\mathbf{Q}}(0)]=0.
\label{spinodal}
\end{equation}
is sufficient for determining the third $\lambda ,$ which turns out to be a
function $\lambda _{\bar{\imath}\bar{\jmath}}(\eta ,x_{1},\zeta )$
independent from the particular closure within the class we are considering.
Since in each matrix element of $\widehat{\mathbf{Q}}$ the quantities
$\bar{y}_{ij}$ and $\tau _{ij}$ appear only in the ratios $\lambda_{ij},$ it 
follows that the angle $\alpha$ [see Eq. (\ref{intro:angle})] will also be
independent of the particular closure.

In the case of a general binary mixture (with two or three
non-vanishing $\epsilon _{ij}$) we expect a dependence of the angle
from the closure, even if this point would deserve further investigation.

\section{The symmetric binary mixture}

\label{sec:symm}

The PY approximation leads, even in the simple binary case, to the solution
of two coupled quartic equations. We then start with a simpler task, akin to
the one already discussed by Chen and Forstmann \cite{Chen92} for a
different potential, of finding the spinodal line and angle $\alpha $
predicted by the mMSA and PY for the symmetric binary mixture. In this
case $x_{1}=x_{2}=1/2$, $\sigma _{1}=\sigma _{2}=\sigma $, and $\epsilon
_{11}=\epsilon _{22}$. By symmetry we must have $\tilde{c}_{11}=\tilde{c}
_{22}$ and from Eq. (\ref{intro:Mc3}) it follows that $\mathbf{M}$ is
diagonal, the cross term $M_{\rho x}$ being identically zero and
\begin{equation}
\lambda _{-}=\min \{M_{\rho \rho },M_{xx}\}~.
\end{equation}
Therefore the symmetric mixture can only have either pure condensation 
$(\alpha =\pm \pi /2)$ or pure demixing $(\alpha =0)$ instabilities.

Moreover for the symmetric mixture we have from
Eqs. (\ref{intro:MS1})-(\ref{intro:MS3})  
\begin{eqnarray}
S_{\rho \rho } &=&\frac{1}{M_{\rho \rho }}~, \\
S_{xx} &=&\frac{1}{4M_{xx}}~, \\
S_{\rho x} &=&0~.
\end{eqnarray}
We see then that on a pure condensation instability $S_{\rho \rho
}(0)\rightarrow \infty $ or $\tilde{h}_{11}(0)+\tilde{h}_{12}(0)\rightarrow
\infty $, whereas on a pure demixing instability $S_{xx}(0)\rightarrow
\infty $ or $\tilde{h}_{11}(0)-\tilde{h}_{12}(0)\rightarrow \infty $, and
each type of instability shows a distinct form of long-range behavior in the
correlation functions.

\subsection{Symmetric mixture in the mMSA}

\label{subsec:symmmsa}

Let us first consider the symmetric mixture within the mMSA. The spinodal
line will be of pure condensation when $M_{\rho \rho }=0$, that is 
\begin{equation}
\tau ^{\ast }=\tau _{\rho }^{\ast }=(1+\nu _{12})\frac{1}{2}\eta \frac{
1-\eta }{1+2\eta }~,  \label{taurho}
\end{equation}
whose maximum in the $(\tau ^{\ast },\eta )$ plane occurs at $\eta =\eta
_{c}^{mMSA}=(\sqrt{3}-1)/2=0.3660\ldots $ (independently of $\nu _{12}$). On
the other hand the spinodal will be a line of pure demixing when $M_{xx}=0$
which has as solution 
\begin{equation}
\tau ^{\ast }=\tau _{x}^{\ast }=(1-\nu _{12})\frac{1}{2}\eta ~.  \label{taux}
\end{equation}

Note that the allowed packing fractions are the ones smaller than the close
packed packing fraction $\eta_0=\pi\sqrt{2}/6=0.7404\ldots$.

For the determinant of $\widehat{\mathbf{Q}}(0)$ we find from Eq. 
(\ref{detQSHS}) 
\begin{equation}
\det [\widehat{\mathbf{Q}}(0)]=\frac{(\tau -\tau _{\rho })(\tau -\tau _{x})}{
\tau ^{2}}\frac{1+2\eta }{(1-\eta )^{2}}~,
\end{equation}
so that the system is unstable when $\tau $ lies between the two roots $\tau
_{\rho }$ and $\tau _{x}$, at a given packing fraction.

\begin{figure}[h!]
\begin{center}
\includegraphics[width=10cm]{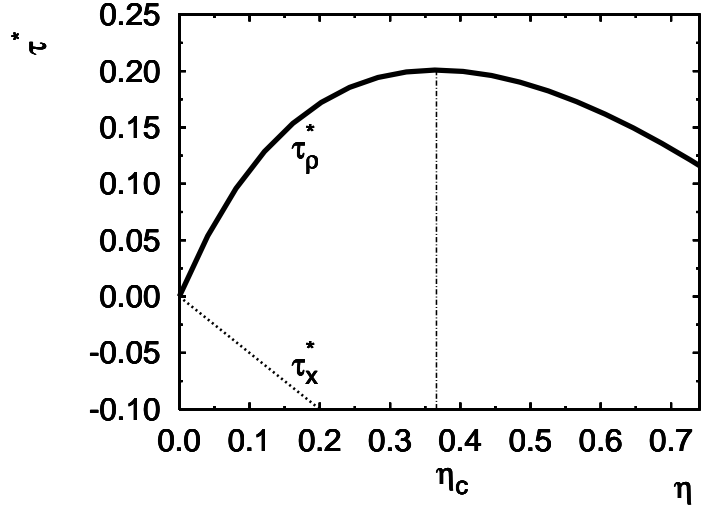}
\end{center}
\caption{Spinodal line (continuous curve) for the symmetric mixture in the
mMSA with $\protect\nu _{12}=2$. The kind of instability is of pure
condensation along the whole spinodal.}
\label{fig:sym1}
\end{figure}

While the condensation line is always present, the existence of a demixing
line depends upon the value of $\nu _{12},$ as expected. When $\nu _{12}\geq
1$ the demixing line $\tau ^{\ast }=\tau _{x}^{\ast }$ lies below the 
$\eta$-axis, and hence the spinodal in the phase diagram $(\tau ^{\ast
},\eta )$ is the curve $\tau ^{\ast }=\tau _{\rho }^{\ast }$ (see
Fig. \ref{fig:sym1}), with the instability being of pure condensation
at all densities.
Notice that this would be the case for Lorentz-Berthelot mixtures for which
we have $\epsilon _{12}\equiv \sqrt{\epsilon _{11}\epsilon _{22}}=\epsilon
_{11}$, which corresponds to $\nu _{12}=1$, that is the one-component case.

\begin{figure}[h!]
\begin{center}
\includegraphics[width=10cm]{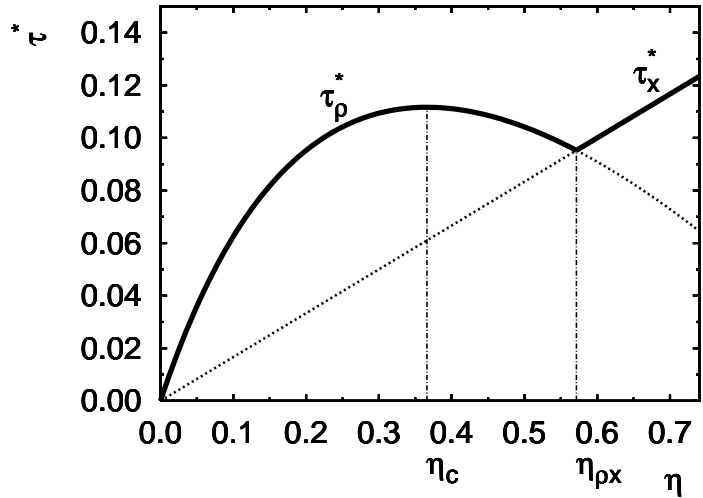}
\end{center}
\caption{Spinodal line (continuous curve) for the symmetric mixture in the
mMSA with $\protect\nu _{12}=2/3$. In this case the instability is of pure
condensation for $\protect\eta <\protect\eta _{\protect\rho x}$ along 
$\protect\tau ^{\ast }=\protect\tau _{\rho}^{\ast }$ and of pure
demixing for $\protect\eta >\protect\eta _{\protect\rho x}$ along 
$\protect\tau ^{\ast }=\protect\tau _{x}^{\ast }$.}
\label{fig:sym2}
\end{figure}

When $\nu _{12}<1$ the two roots $\tau ^{\ast }=\tau _{\rho }^{\ast }$
and $\tau ^{\ast }=\tau _{x}^{\ast }$ intercept at a point
\cite{Wilding98} (see Fig. \ref{fig:sym2}) having packing fraction 
\begin{equation}
\eta =\eta _{\rho x}=\frac{2\nu _{12}}{3-\nu _{12}}<1~,
\end{equation}
so the instability is of pure condensation for $\eta <\eta _{\rho x}$ and of
pure demixing for $\eta >\eta _{\rho x}$. 

\subsection{Symmetric mixture in the PY}

\label{subsec:sympy}

In the PY approximation we first need to determine the cavity functions at
contact. Eqs. (\ref{PYyij}) for the binary symmetric mixture can be recast
into the following form 
\begin{eqnarray} \label{perram-sym-11}
\lambda _{11}\tau _{11}-\frac{1}{2}\eta \left( \frac{1}{12}\lambda _{11}^{2}-
\frac{1}{\Delta }\lambda _{11}\right) &=&\bar{y}_{11}^{HS}+\frac{1}{2}\eta
\left( \frac{1}{12}\lambda _{12}^{2}-\frac{1}{\Delta }\lambda _{12}\right) ~,
\\ \label{perram-sym-12}
\lambda _{12}\tau _{12}\left[ 1+\frac{1}{\tau _{12}}\left( \frac{\eta }{
2\Delta }-\frac{1}{12}\eta \lambda _{11}\right) \right] &=&\bar{y}_{12}^{HS}-
\frac{\eta }{2\Delta }\lambda _{11}~,
\end{eqnarray}
where 
\begin{equation}
\bar{y}_{11}^{HS}=\bar{y}_{12}^{HS}=\bar{y}^{HS}=\frac{2+\eta }{2(1-\eta
)^{2}}~,
\end{equation}
is the HS expression for the cavity functions at contact. Substitution of
Eq. (\ref{perram-sym-12}) into Eq. (\ref{perram-sym-11}) leads to a 
quartic equation for $\lambda _{11}$. The solution for the cavity
functions at contact can then be written as 
\begin{eqnarray}
\frac{\bar{y}_{11}}{\tau _{11}} &=&R~, \\
\frac{\bar{y}_{12}}{\tau _{12}} &=&\frac{\bar{y}^{HS}-\frac{\eta }{2\Delta }R
}{\tau _{12}\left[ 1+\frac{1}{\tau _{12}}\left( \frac{\eta }{2\Delta }-\frac{
1}{12}\eta R\right) \right] }~,
\end{eqnarray}
where $R$ is a solution of the quartic equation.

In order to find the physically meaningful zeroes of $M_{\rho \rho }$
and $M_{xx}$ we proceed as follows. First we compute all the four
roots $R_{i}$, $i=a,b,c,d$ of the quartic equation and hence
$(\bar{y}_{11})_{i}=(\bar{y}_{11})_{i}(\tau ^{\ast },\eta ,\nu
_{12})$, and $(\bar{y}_{12})_{i}=(\bar{y}_{12})_{i}(\tau ^{\ast },\eta
,\nu _{12})$ are the cavity functions at contact obtained using the
root $R_{i}$, while $(M_{\rho \rho })_{i}$ and $(M_{xx})_{i}$ are the
diagonal elements of $\mathbf{M}$ obtained using for the cavity
functions at contact $(\bar{y}_{11})_{i}$ and $(\bar{y}_{12})_{i}$. As
it turns out, only two roots $R_{i}$ will give physically admissible 
cavity functions at contact. Then we compute the zeroes of $(M_{\rho \rho
})_{i}$, denoted as $\tau ^{\ast }=(\tau _{\rho }^{\ast })_{i}(\eta ,\nu
_{12})$, and of $(M_{xx})_{i}$, denoted as $\tau ^{\ast }=(\tau _{x}^{\ast
})_{i}(\eta ,\nu _{12})$. Then physical zeroes are then selected by the
requirement that 
\begin{equation}
\lim_{\nu _{12}\rightarrow 1}(\bar{y}_{11})_{i}((\tau _{\alpha }^{\ast
})_{i},\eta ,\nu _{12})=\lim_{\nu _{12}\rightarrow 1}(\bar{y}
_{12})_{i}((\tau _{\alpha }^{\ast })_{i},\eta ,\nu _{12})=\bar{y}
_{+}^{oc}((\tau _{\alpha }^{\ast })_{i},\eta )~~~\alpha =\rho ,x~,
\label{pysymm:y-phys}
\end{equation}
where $\bar{y}_{+}^{oc}$ is the physical cavity function at contact for the
one-component system 
\begin{equation}
\bar{y}_{\pm }^{oc}(\tau ,\eta )=\frac{\bar{y}^{HS}}{\frac{1}{2}\left[ 1+
\frac{\eta }{\Delta }\frac{1}{\tau }\pm \sqrt{\left( 1+\frac{\eta }{\Delta }
\frac{1}{\tau }\right) ^{2}-\frac{\eta }{3}\bar{y}^{HS}\frac{1}{\tau ^{2}}}
\right] }~.
\end{equation}
Using constraint (\ref{pysymm:y-phys}) we find that the correct physical
solution is $R=R_{b}$ at high $\eta $, the only one such that 
\begin{eqnarray}
\lim_{\tau \rightarrow \infty }(\bar{y}_{11})_{b} &=&\lim_{\tau \rightarrow
\infty }(\bar{y}_{12})_{b}=\bar{y}^{HS}~, \\
\lim_{\eta \rightarrow 0}(\bar{y}_{11})_{b} &=&\lim_{\eta \rightarrow 0}(
\bar{y}_{12})_{b}=1~,
\end{eqnarray}
while at small $\eta $ the solution to use is $R=R_{a}$ such that
condition (\ref{pysymm:y-phys}) is satisfied. As for the one-component
system there is an interval $[0,\eta _{e}]$ where there are no
physical zeroes. For the one-component case the spinodal 
\begin{equation}
\tau ^{\ast }=\tau _{oc}^{\ast }=\frac{1+4\eta -14\eta ^{2}}{12(1-\eta
)(1+2\eta )}~,  \label{py-oc-spinodal}
\end{equation}
exists only if $\eta >\eta _{e}$ where $\eta _{e}=\eta _{c}^{PY}=(3\sqrt{2}
-4)/2=0.1213\ldots $ and $\eta _{c}^{PY}$ is the PY critical packing
fraction. For the binary symmetric mixture, numerical results strongly
suggests the coincidence of $\eta _{e}$ with the critical packing fraction
(see Figs. \ref{fig:py-sym1}, \ref{fig:py-sym1-oclim},
and \ref{fig:py-sym1-oclim}) but we have not succeeded in proving it (nor in
determining an expression for it). The unphysical continuation of the pure
condensation spinodal in the range $[0,\eta _{e}]$ is given by the
root $R=R_{c}$ such that 
\begin{equation}
\lim_{\nu _{12}\rightarrow 1}(\bar{y}_{11})_{c}((\tau _{\alpha }^{\ast
})_{c},\eta ,\nu _{12})=\lim_{\nu _{12}\rightarrow 1}(\bar{y}
_{12})_{c}((\tau _{\alpha }^{\ast })_{c},\eta ,\nu _{12})=\bar{y}
_{-}^{oc}((\tau _{\alpha }^{\ast })_{c},\eta )~~~\alpha =\rho ,x~.
\label{pysymm:y-unphys}
\end{equation}
Notice that this solution would also give, in the same range of $\eta $, an
unphysical spinodal of pure demixing whenever $\nu _{12}<1$.

The zeroes $\tau ^{\ast }=(\tau _{\rho }^{\ast })_{i}$ and $\tau ^{\ast
}=(\tau _{x}^{\ast })_{i}$ are shown in Fig. \ref{fig:py-sym1} for
$\nu_{12}=2,$ and for $\nu_{12}=2/3$ (the same 
conditions as in Figs. \ref{fig:sym1} and \ref{fig:sym2}, respectively). As
it happened in the mMSA, for $\nu _{12}>1$ there is only a spinodal of pure
condensation, while for $\nu _{12}<1$ a spinodal of pure demixing appears at
high $\eta ,$ as expected on physical grounds. However, unlike the mMSA
case, the pure demixing and the pure condensation lines do not merge. Also
the shapes and numerical values of the PY spinodals significantly differ
from the mMSA ones. In Fig. \ref{fig:py-sym1-oclim} we select $\nu
_{12}$ slightly above 1 and slightly 
below 1 in order to check the correct convergence towards the
one-component case. At $\nu _{12}=1/1.1$ the line of pure demixing appears
in the physically non-accessible region $\eta >\eta _{0}$.

\section{Two paradigmatic systems}

\label{sec:mMSA-PY}

The next two mixtures can be regarded as paradigmatic examples of a system
where one expects to have a predominant condensation or predominant demixing
type of thermodynamic instability. The thermodynamics of these mixtures have
been previously investigated by Barboy and Tenne \cite{Barboy79} within the
PY approximation. In the following we shall extend this analysis of the
instability type both within mMSA and PY closures. The two systems are
defined as follows: (System A) $\epsilon _{12}>0$, $\epsilon _{11}=\epsilon
_{22}=0$; (System B) $\epsilon _{11}>0$, $\epsilon _{12}=\epsilon _{22}=0$.
System A corresponds to a fluid where the HS potential acts between like
particles and the SHS potential between unlike particles, while System B
corresponds to HS (species 2) in a SHS fluid (species 1).
Alternatively, on regarding the large spheres as the ``solute'' and
the small spheres as the ``solvent'', Systems A and B can be reckoned
as a schematic model mimicking a ``good'' and a ``poor'' solvent,
respectively \cite{Penders91}.  

For System A we have $\lambda _{11}^{BT}=\lambda _{22}^{BT}=0$, so
Eq. (\ref{PYyij}) reduces to a a linear equation for $\bar{y}_{12}$
with the following solution (which corrects Eq. (64) of Barboy and Tenne) 
\begin{equation}
\bar{y}_{12}=\frac{\bar{y}_{12}^{HS}}{1+\frac{\xi _{2}}{2\Delta }\frac{
\sigma _{12}}{\tau _{12}}}~,
\end{equation}
where 
\begin{equation}
\bar{y}_{12}^{HS}=\frac{1}{\Delta }+\frac{3}{2}\frac{\xi _{2}}{\Delta ^{2}}
\frac{\sigma _{1}\sigma _{2}}{\sigma _{12}}~,
\end{equation}
is the HS expression for the contact cavity function.

For System B we have $\lambda _{12}^{BT}=\lambda _{22}^{BT}=0$, so
Eq. (\ref{PYyij}) reduces to a quadratic equation for $\bar{y}_{11}$. The only
solution which reduces to the HS expression for $\tau _{11}\rightarrow
\infty $, is (identical to Eq. (57) of Barboy and Tenne) 
\begin{equation}
\bar{y}_{11}=\frac{\bar{y}_{11}^{HS}}{\frac{1}{2}\left[ 1+\frac{\eta _{1}}{
\Delta }\frac{1}{\tau _{11}}+\sqrt{\left( 1+\frac{\eta _{1}}{\Delta }\frac{1
}{\tau _{11}}\right) ^{2}-\frac{\eta _{1}}{3}\bar{y}_{11}^{HS}\frac{1}{\tau
_{11}^{2}}}\right] }~,
\end{equation}
where 
\begin{equation}
\bar{y}_{11}^{HS}=\frac{1}{\Delta }+\frac{3}{2}\frac{\xi _{2}}{\Delta ^{2}}
\sigma _{1}~,
\end{equation}
is the HS expression for the contact cavity function. The instability lines
are again given by Eq. (\ref{spinodal}).

Let $\lambda _{\bar{\imath}\bar{\jmath}}(\eta ,x_{1},\zeta )$ be the
solution of the spinodal equation (\ref{spinodal}) for the only
non-vanishing $\lambda _{ij}$. As the cavity functions must be positive, the
spinodal exists only for those values of $\eta ,x,\zeta $ for which $\lambda
_{\bar{\imath}\bar{\jmath}}>0$. It may also happen (and it does in the PY
case) that the spinodal equation 
\begin{equation}
\lambda _{\bar{\imath}\bar{\jmath}}^{(closure)}=\frac{\bar{y}_{\bar{\imath}
\bar{\jmath}}^{(closure)}(\tau _{\bar{\imath}\bar{\jmath}},\eta ,x_{1},\zeta
)}{\tau _{\bar{\imath}\bar{\jmath}}}=\lambda _{\bar{\imath}\bar{\jmath}
}(\eta ,x_{1},\zeta )~,
\end{equation}
upon choosing the correct physical solution for $\bar{y}_{\bar{\imath}\bar{
\jmath}}^{(closure)}$, does not have any real positive solutions for $\tau $
, at certain values of $\eta ,x,\zeta $. For these values the spinodal
predicted by the particular closure has loss of solution and the predicted
value for the angle $\alpha $ has clearly no physical meaning.

\subsection{Instabilities for System A}

On setting 
\begin{equation}
\lambda _{A}^{BT}=3+\sqrt{\left( 3+\frac{\Delta }{\eta _{1}}\right) \left( 3+
\frac{\Delta }{\eta _{2}}\right) }~,
\end{equation}
for System A the solution of Eq. (\ref{spinodal}) within the mMSA
approximation is 
\begin{equation}
\tau _{12}^{mMSA}=\frac{\Delta }{\lambda _{A}^{BT}}\frac{\sigma _{12}^{2}}{
\sigma _{1}\sigma _{2}}~.  \label{A:spinodal-mMSA}
\end{equation}
while in the PY is 
\begin{equation}
\tau _{12}^{PY}=\bar{y}_{12}^{HS}\tau _{12}^{mMSA}-\frac{\xi _{2}\sigma _{12}
}{2\Delta }~,  \label{A:spinodal-PY}
\end{equation}
and, in the limit of high dilution while keeping $\tau _{12}^{mMSA}$
constant, one finds $\tau _{12}^{PY}\rightarrow \tau _{12}^{mMSA},$ as
expected in view of the fact that the PY contact cavity functions converge
towards the mMSA contact cavity functions.

In order to exist, the instability line must clearly lie on the $\tau
_{12}>0 $ side of the $(\tau _{12},\eta )$ plane. It is easy to see that,
while 
\begin{equation}
\left. \frac{d\tau _{12}^{mMSA}}{d\eta }\right\vert _{\eta =0}>0~~
\mbox{for any
choice of $x_1$ and $\zeta$}~,  \label{mMSAdtdr}
\end{equation}
we have that ($\sigma _{1}<\sigma _{2}$) 
\begin{equation}
\left. \frac{d\tau _{12}^{PY}}{d\eta }\right\vert _{\eta =0}>0~~
\mbox{only
when}~~\frac{\sigma _{2}}{\sigma _{1}+\sigma _{2}}<x_{1}<\frac{\sigma
_{2}^{3}}{\sigma _{1}^{3}+\sigma _{2}^{3}}~.  \label{PYdtdr}
\end{equation}
So in the PY approximation the thermodynamic instability disappears as 
$x_{1} $ falls outside the range indicated in Eq. (\ref{PYdtdr}).

In Figs. \ref{fig:system-a-mmsa-t} and \ref{fig:system-a-py-t} we depict the
mMSA and PY spinodals, respectively, at a given value of $\zeta $ and three
different values of $x_{1}$ for which the PY spinodal does exist. One
clearly sees that conditions (\ref{mMSAdtdr}) and (\ref{PYdtdr}) result in a
large scale difference between the two plots. 

As regards the angle $\alpha $, we know, from the discussion at the end of
Section \ref{sec:SHSmixture}, that the angles predicted by the two
approximations are the same, and from Fig. \ref{fig:system-a-a} we see that
the kind of instability is mainly of \ condensation type in accord with what
we expected from the outset. Two exact limits are worth mentioning. First,
the infinite dilute limit 
\begin{equation}
\lim_{\eta \rightarrow 0}\alpha =\arctan \left( \frac{\sqrt{x_{2}/x_{1}}+
\sqrt{\sigma _{2}/\sigma _{1}}}{\sqrt{\sigma _{2}x_{2}/(\sigma _{1}x_{1})}-1}
\right) ~,
\end{equation}
provides an analytical check of the numerical results reported in Fig.
\ref{fig:system-a-a}. Second, when $\sigma _{2}\gg \sigma _{1}$ one
obtains 
\begin{equation}
\lim_{\zeta \rightarrow \infty }\alpha =\arctan \sqrt{x_{1}/x_{2}}~.
\label{alpha-zeta-limit}
\end{equation}
This result bears an interesting physical interpretation. As the
fraction $x_{2}$ of large particles decreases, the angle $\alpha $
tends to $\pi /2$, 
that is to a condensation instability. This is in striking contrast
with what one would expect for HS on the basis of an entropic depletion
mechanism \cite{Asakura54,Hansen-Barrat}, which would tend to favor
demixing in a system with a small number of large spheres. The reason for
this can be traced back to the fact that in System A unlike particles have
attractive interactions, thus preventing smaller particles to slip out from
the interstitial region between two larger spheres. 
This interpretation also holds true if one regards system A as a good
solvent.

\subsection{Instabilities for System B}

Denoting 
\begin{equation}
\lambda _{B}^{BT}=\frac{(1+2\eta )(1-\eta )}{\eta _{1}(1-\eta )+3\eta
_{1}\eta _{2}}~,
\end{equation}
for System B the solution of Eq. (\ref{spinodal}) within the mMSA is 
\begin{equation}
\tau _{11}^{mMSA}=\frac{\Delta }{\lambda _{B}^{BT}}~,
\label{B:spinodal-mMSA}
\end{equation}
while in the PY approximation is 
\begin{equation}
\tau _{11}^{PY}=\bar{y}_{11}^{HS}\tau _{11}^{mMSA}+\frac{\eta _{1}}{12\tau
_{11}^{mMSA}}-\frac{\eta _{1}}{\Delta }~,  \label{B:spinodal-PY}
\end{equation}
for 
\begin{equation}
\tau _{11}^{PY}>\frac{\eta _{1}}{\Delta }\left( \frac{\lambda _{B}^{BT}}{6}
-1\right) ~.  \label{sysB:PY-sp-ex}
\end{equation}
In view of the above constraint, there is an interval $\eta \in \lbrack
0,\eta _{e}]$ where no physical spinodal exists. We stress that only for the
one-component SHS limit ($x_{2}=0$) one finds that $\eta _{e}=\eta _{c}$,
with $\eta _{c}$ being the critical packing fraction, whereas in the more
general case, studied here, this occurrence is no longer true, as shown in
Fig. \ref{fig:system-b-py-t}. Once again $\tau _{11}^{PY},$ as given
in Eq. (\ref{B:spinodal-PY}), reduces to $\tau _{11}^{mMSA}$, in the
limit of high dilution, with $\tau _{11}^{mMSA}$ kept constant.
However, unlike $\tau _{11}^{mMSA},$ which is always a concave function of $\eta $
for any choice of $x_{1}$ and $\zeta $, $\tau _{11}^{PY}$, it may display a van der Waals loop (see
Fig. \ref{fig:system-b-py-t}) as a function of $\eta$. 
The shape of the spinodal is strongly dependent on the content of the HS
component in the mixture. When $x_{1}<\bar{x}_{1}$ ($\bar{x}_{1}\approx
0.8681\ldots $ when $\zeta =1$) the spinodal is a monotonously increasing
function of $\eta $, while for $x_{1}>\bar{x}_{1}$ a loop appears. This
point has already been emphasized by Barboy and Tenne \cite{Barboy79}.

As previously remarked, even in this case both mMSA and PY results for 
$\alpha $ coincide in the respective range of existence. In Fig. 
\ref{fig:system-b-py-a-y1} we see that the
instability for System B tends to pure demixing for $\zeta =1$ and
large $\eta $. As $\zeta $ is increased, one finds the same limit 
(\ref{alpha-zeta-limit}) as for System A. Once again the osmotic depletion
mechanism fails because of the presence of stickiness this time among the
small particles. As a further support to this interpretation, one also finds
in the opposite limit 
\begin{equation}
\lim_{\zeta \rightarrow 0}\alpha =\arctan \left( \frac{x_{1}-\eta }{\sqrt{
x_{1}x_{2}}}\right) ~.  \label{B-od}
\end{equation}
In this case, when $x_{1}=\eta$ the instability of the system is of a
pure demixing type, so the solvent (particles of species 2) is a
``poor'' one. This is because the smaller particles (species 2)
interact as HS both with larger spheres (species 1) and with each other.
Hence, not only the depletion mechanism is not opposed in the present case,
but, quite on the contrary, is favored by the attraction occurring between
two big spheres (see Fig. \ref{fig:od}). 
This results into the possibility for the existence of a demixing instability even 
if the HS binary mixture, within the
closures considered here, does not have any instability [see Eq. 
(\ref{detQHS})]. One can also show that 
\begin{equation}
\lim_{\eta \rightarrow 0}\alpha ^{mMSA}=\arctan \sqrt{x_{1}/x_{2}}~,
\end{equation}
where $\alpha ^{mMSA}$ is the angle predicted by the mMSA, whose spinodal
does not have loss of solution at small $\eta $, or, upon using $\eta
_{1},\eta _{2},y$ as independent variables, $\lim_{\eta _{1}\rightarrow
0}\alpha ^{mMSA}=0$. 

Before closing this section, a word of caution should be given on the
aforementioned interpretations. In order to have a clear and
quantitative understanding of the depletion mechanism discussed in
this section (for both System A and B), the depletion potential, that
is the effective potential among the large spheres mediated by the
presence of the small ones, should be computed. Hence, the
aforementioned scenarios should only be considered as a plausible
possibility rather than a definite statement.  

\section{Five binary mixtures treated with mMSA}

\label{subsec:cases}

As a final point it is instructive to consider a more general example. To
this aim, it proves convenient to relate the adhesion strengths $\epsilon
_{ij}$ to the particle sizes $\{\sigma _{i}\}$. Our past experience 
\cite{Fantoni05} suggests to consider five different cases, obtained
setting $\epsilon _{ij}/\epsilon _{0}=\mathcal{F}_{ij}^{\mu}
(\sigma_{1},\sigma_{2})$ for $\mu =1$, $2$, $3$, $4$, and $5$. The
functions $\mathcal{F}^{\mu}$ are selected as follows \cite{Fantoni05} 
\begin{equation}
\frac{\epsilon _{ij}}{\epsilon _{0}}=\left\{ 
\begin{array}{ll}
\langle \sigma \rangle ^{2}/\sigma _{ij}^{2} & \mbox{Case I}~, \\ 
\sigma _{i}\sigma _{j}/\sigma _{ij}^{2} & \mbox{Case II}~, \\ 
\langle \sigma ^{2}\rangle /\sigma _{ij}^{2} & \mbox{Case III}~, \\ 
1 & \mbox{Case IV}~, \\ 
\langle \sigma \rangle /\sigma _{ij} & \mbox{Case V}~,
\end{array}
\right.   \label{Cases}
\end{equation}
where $\langle F\rangle \equiv \sum_{i}x_{i}F_{i}$. A critical
justification leading to the above choice can be found in
Ref. \cite{Fantoni05}. Note that since for all five Cases the
$\epsilon_{ij}$ are homogeneous functions of order zero in the
diameters $\{\sigma_i\}$, the corresponding mixtures are invariant
under a transformation where $V\to\lambda V$ and all
$\sigma_i\to\lambda\sigma_i$ with $\lambda$ a scale factor \cite{Sollich05}. 

We have calculated the angle $\alpha$ defined in Eq. (\ref{intro:angle}) on
the spinodal [Eq. \ref{spinodal}] for all the Cases listed in (\ref{Cases})
within the mMSA closure. The angle $\alpha$ turns out to be the same for
Case I and III. The results are shown in Figs. \ref{fig:Cases1} and 
\ref{fig:Cases2} for $x_1=1/2$ and two different values of $\zeta$. We
have only 
considered packing fractions $\eta\le\eta_m=\pi\sqrt{2}/6$, where $\eta_m$
is the maximum packing fraction for a ``completely demixed'' HS mixture 
(i.e. the packing fraction of a mixture where the spheres of species 1
are in a close packed configuration occupying a volume $V_1$ and the spheres
of species 2 are in a closed packed configuration occupying a volume $V_2$
such that $V_2\cap V_1=0$). It gives a lower bound to the true maximum
packing fraction. 

In Cases I and III we have pure condensation as $\eta \rightarrow 0$. Case V
display a pure condensation point at small but non-zero values of $\eta $.
In Case II we find a pure demixing point at high $\eta $, for sufficiently
large $\zeta $ in the same region where in Case IV we have a pure
condensation point. The packing fraction of pure demixing for Case II can be
easily calculated to be 
\begin{equation}
\eta =\frac{\langle \sigma \rangle \langle \sigma ^{3}\rangle }{\langle
\sigma ^{4}\rangle }~,
\end{equation}
which turns out to be very close, albeit in general not coincident, with the
packing fraction at which we find pure condensation in Case IV.

We remark that (both for mMSA and PY) the presence of an instability curve
for the SHS model is entirely due to the stickiness, since in the HS
limit ($\tau \rightarrow \infty $) we have 
\begin{equation}
\lim_{\tau \rightarrow \infty }\det [\widehat{\mathbf{Q}}(0)]=\frac{1+2\eta 
}{(1-\eta )^{2}}~,  \label{detQHS}
\end{equation}
which is always a positive quantity. Eq. (\ref{detQHS}) can be derived from
Eq. (\ref{detQSHS}) by noticing that the contact values of the partial
cavity functions $\bar{y}_{ij}$ must remain finite as $\tau \rightarrow
\infty $. So the above statement is actually valid for any closure in which
the partial direct correlation functions vanish beyond $\sigma _{ij}$. In
particular it is valid for the mMSA and the PY \cite{Lebowitz64}
approximations. For other, thermodynamically more consistent closures, the
statement is no longer true since phase separation has been observed for
highly asymmetric HS binary mixtures \cite{Biben91}.

\section{Conclusions}

In this work we have applied the method devised by Chen and Forstmann 
\cite{Chen92} to characterize the kind of thermodynamic instability to
a number of carefully selected SHS binary systems. The crucial
quantity turns out to 
be the Chen and Forstmann angle $\alpha $, see Eq. (\ref{intro:angle}), on
the spinodal: when $\alpha $ is close to $0$ the instability is of the pure
demixing type, whereas a value close to $\pm \pi /2$ indicates a pure
condensation instability.

The presence of adhesion between the spheres results in the existence of
thermodynamic instabilities for the SHS model when treated within
closures having the direct correlation functions vanishing beyond the hard
core ranges, whereas it is known that the HS mixture within the same 
approximations do not show any instability [see Eq. (\ref{detQHS})].

We have first considered the symmetric binary mixture in the mMSA (see
Section \ref{subsec:symmmsa}) and in the PY approximation (see Section 
\ref{subsec:sympy}). This latter case was already considered by Chen and Forstmann for
a different potential. We have found that when $\epsilon _{11}\leq \epsilon
_{12}$ the instability is of pure condensation along the whole spinodal [see
Fig. \ref{fig:sym1} and Eq. (\ref{taurho}) for the mMSA, and Figs. 
\ref{fig:py-sym1} and \ref{fig:py-sym1-oclim} for the PY], while when $\epsilon
_{11}>\epsilon _{12}$ a pure demixing spinodal appears at large packing
fractions [see Fig. \ref{fig:sym2} and Eqs. (\ref{taurho}) and (\ref{taux})
for the mMSA, and Figs. \ref{fig:py-sym1} and \ref{fig:py-sym1-oclim} for
the PY], all within their respective limits of validity. This general
behavior appears to be characteristic of symmetric binary mixtures, in the
sense that it is observed in systems with pair potentials more
\textquotedblleft complex\textquotedblright\ than the SHS potential (hard
spheres with Yukawa tails \cite{Jedrzejek87}, square well \cite{Wilding98},
Lennard-Jones \cite{Antonevych02}, etc.) which do not admit analytic
solutions. The condensation and demixing lines are found to meet at a point
in the mMSA, whereas they do not merge within the PY approximation.

Other two interesting examples can be treated in detail from an analytical
point of view as discussed in Section \ref{sec:mMSA-PY}. We compared the
spinodals and the angles $\alpha $ predicted by mMSA with those predicted by
PY for a binary mixture with $\epsilon _{12}>0$ and $\epsilon _{11}=\epsilon
_{22}=0$ (System A) and one with $\epsilon _{11}>0$ and $\epsilon
_{12}=\epsilon _{22}=0$ (System B). Being the SHS interaction attractive,
one should expect System A to present mainly condensation instabilities and
System B mainly demixing instabilities. These choices for the $\epsilon _{ij}
$ reduce the equations (\ref{PYyij}) for the contact values of the cavity
functions in the PY approximation at most to a quadratic one, simplifying
calculations considerably. We find that the spinodals predicted by the two
approximations are very different both quantitatively and qualitatively [see
Figs. \ref{fig:system-a-mmsa-t} and \ref{fig:system-a-py-t}, and Eqs. 
(\ref{A:spinodal-mMSA}) and (\ref{A:spinodal-PY}) for System A, and
Fig. \ref{fig:system-b-py-t} and Eqs. (\ref{B:spinodal-mMSA}) and
(\ref{B:spinodal-PY}) for System B]. Nonetheless the corresponding
angles $\alpha $ do not 
depend on the closure, when this is chosen within the GPY large class
containing mMSA and PY as particular cases. In agreement with our
expectations, we find that the instabilities of System A are predominantly
of the condensation type (see Fig. \ref{fig:system-a-a}), while the ones of
System B of the demixing type when $\zeta \simeq 1$ (see Fig. 
\ref{fig:system-b-py-a-y1}). For System
B when we have a small number of large spheres of species 1, the
demixing instability may be favored by both the osmotic depletion
mechanism \cite{Asakura54} and the stickiness between the large
spheres (see Fig. \ref{fig:od}).  

In the more general case, the pair potential depends in general on 3
parameters: the ratio of the sphere diameters of the two species, $\zeta
=\sigma _{2}/\sigma _{1,}$ and two dimensionless parameters which measure
the relative strength of surface adhesiveness, $\nu _{22}=\epsilon
_{22}/\epsilon _{11}$ and $\nu _{12}=\epsilon _{12}/\epsilon _{11}$. A
reduction occurs when the latters are connected to the former through
plausible relationships $\epsilon _{ij}=\epsilon _{0}\mathcal{F}_{ij}(\sigma
_{1},\sigma _{2})$. Following our previous work \cite{Fantoni05}, we have
considered five possible cases [see Section \ref{subsec:cases} and
Eq. (\ref{Cases})]. We find that four of the five cases exhibit very
distinct types 
of instabilities (see Figs. \ref{fig:Cases1} and \ref{fig:Cases2}): Cases I
and III have the same angle $\alpha $, with pure condensation at $\eta
\rightarrow 0$ and predominant demixing for $\eta >0$; Case V has a pure
condensation instability point at low packing fractions; Case IV has a pure
condensation instability point at high packing fractions provided that 
$\zeta $ is sufficiently large, whereas Case II has a pure demixing
instability point under the same conditions.

It would be desirable to extend the present study in two
respects. First it would be interesting to consider different, more
sophisticated, closures, in view of
our results on the two examples (denoted as System A and System B) where the
angle $\alpha $ is shown to be independent of the particular closure within
the GPY class, in spite of a large difference in the corresponding
instability curves. Second, it would be nice to test the analytical predictions
given in this work against numerical simulations, with a particular attention
to what concerns the depletion mechanism. We plan to address both issues in a
future work.

\appendix

\section{Thermodynamic relations for the elements of the $M$ matrix}
\label{app1}

In this appendix we gather together some well known relationships between
thermodynamic quantities and the results obtained in the main text. The
Ashcroft-Langreth partial structure factors \cite{Ashcroft67} of an
homogeneous and isotropic $p-$component mixture are related to the partial
total correlation functions as 
\begin{equation}
S_{ij}(k)=\delta _{ij}+\rho \;\sqrt{x_{i}x_{j}}\tilde{h}_{ij}(k)~,
\end{equation}
where $x_{i}=\langle N_{i}\rangle /\langle N\rangle $ is the molar fraction
of particles of species $i$ and $\rho $ the total density of the mixture.
From the normalization condition for the partial pair distribution functions
of the grand canonical ensemble follows 
\begin{equation}
S_{ij}(0)=\sqrt{\frac{x_{i}}{x_{j}}}\left( \frac{\langle N_{i}N_{j}\rangle
-\langle N_{i}\rangle \langle N_{j}\rangle }{\langle N_{i}\rangle }\right) ~,
\end{equation}
The matrix $\mathbf{\tilde{A}}$, defined in Eq. (\ref{intro:tildeA}) of the
text, is related to the structure factors by 
\begin{equation}
S_{ij}(k)=[\mathbf{\tilde{A}}^{-1}]_{ij}(k)~.
\end{equation}

We now relate composition fluctuations to thermodynamic quantities. The
grand partition function is 
\begin{equation}
e^{-\beta \Omega }=\sum_{N_{1},\ldots ,N_{p}=0}^{\infty }e^{\beta \lbrack
\sum_{i=1}^{p}N_{i}\mu _{i}-A(T,V,\{N_{i}\})]}~,
\end{equation}
where $A(T,V,\{N_{i}\})$ is the Helmholtz free energy of a member of the
grand canonical ensemble with given number of particles of each species, and
the chemical potentials $\{\mu _{i}\}$ are to be determined from the average
number of particles of each species 
\begin{equation}
\langle N_{i}\rangle =\sum_{N_{1},\ldots ,N_{p}=0}^{\infty }N_{i}e^{\beta
\lbrack \Omega +\sum_{i=1}^{p}N_{i}\mu _{i}-A(T,V,\{N_{i}\})]}~.
\end{equation}
We immediately find 
\begin{equation}
\left( \frac{\partial \Omega }{\partial \mu _{i}}\right) _{T,V,\{\mu _{\bar{
\imath}}\}}=-\langle N_{i}\rangle ~,
\end{equation}
and 
\begin{eqnarray}
\frac{1}{\beta }\left( \frac{\partial N_{i}}{\partial \mu _{j}}\right)
_{T,V,\{\mu _{\bar{\jmath}}\}} &=&\langle N_{i}\rangle \left( \frac{\partial
\Omega }{\partial \mu _{j}}\right) _{T,V,\{\mu _{\bar{\jmath}}\}}+\langle
N_{i}N_{j}\rangle   \notag \\
&=&\langle N_{i}N_{j}\rangle -\langle N_{i}\rangle \langle N_{j}\rangle =
\sqrt{x_{i}x_{j}}S_{ij}(0)\langle N\rangle ~,
\end{eqnarray}
where the index $\bar{\imath}$ denotes all species different from $i$. Since
the thermodynamic derivatives $(\partial N_{i}/\partial \mu _{j})_{T,V,\{\mu
_{\bar{\jmath}}\}}$ are the elements of the inverse of the matrix whose
elements are $(\partial \mu _{i}/\partial N_{j})_{T,V,\{N_{\bar{\jmath}}\}}$
we can invert the above relation to read 
\begin{eqnarray}
\beta \left( \frac{\partial \mu _{i}}{\partial N_{j}}\right) _{T,V,\{N_{\bar{
\jmath}}\}} &=&\frac{1}{\langle N\rangle \sqrt{x_{i}x_{j}}}[\mathbf{S}
^{-1}]_{ij}(0)  \notag  \label{app1:dmdn} \\
&=&\frac{1}{V\rho \sqrt{x_{i}x_{j}}}\tilde{A}_{ij}(0)~,
\end{eqnarray}
where we indicated with $\mathbf{S}$ the matrix whose elements are the
partial structure factors.

We now define the partial volumes as 
\begin{equation}
v_{i}=\left( \frac{\partial V}{\partial N_{i}}\right) _{T,P,\{N_{\bar{\imath}
}\}}~.
\end{equation}
Since the total volume is an homogeneous function of order one in the
extensive variables we must have 
\begin{equation}
\sum_{i=1}^{p}N_{i}v_{i}=V~,  \label{app1:V}
\end{equation}
since the Gibbs free energy $G=G(T,P,\{N_{i}\})$ is an homogeneous function
of order one in the extensive variables we must have 
\begin{equation}
\sum_{i=1}^{p}N_{i}\mu _{i}=G~,  \label{app1:G}
\end{equation}
so in particular the chemical potentials will be homogeneous functions of
order zero in the variables $\{N_{i}\}$, we can then write $\mu _{i}=\mu
_{i}(T,P,\{\{N_{i}\}\})$ where with the symbol $\{\{N_{i}\}\}$ we mean that
the variables $\{N_{i}\}$ can appear only as ratios. We also find 
\begin{equation}
\left( \frac{\partial \mu _{i}}{\partial N_{j}}\right) _{T,V,\{N_{\bar{\jmath
}}\}}=\left( \frac{\partial \mu _{i}}{\partial N_{j}}\right) _{T,P,\{\{N_{
\bar{\jmath}}\}\}}+\frac{v_{i}v_{j}}{V\chi _{T}}~,  \label{app1:dmudN}
\end{equation}
where $\chi _{T}$ is the isothermal compressibility 
\begin{equation}
\chi _{T}=-\frac{1}{V}\left( \frac{\partial V}{\partial P}\right)
_{T,\{N_{i}\}}~.
\end{equation}
Notice also that taking the partial derivative of Eq. (\ref{app1:G}) with
respect to $N_{j}$ at constant $T$, $P$, and $\{N_{\bar{\jmath}}\}$ we find
the following Gibbs-Duhem relation 
\begin{equation}
\sum_{i=1}^{p}N_{i}\left( \frac{\partial \mu _{i}}{\partial N_{j}}\right)
_{T,P,\{\{N_{\bar{\jmath}}\}\}}=0~.  \label{app1:G-D}
\end{equation}

We want now find thermodynamic relations for the matrix elements $M_{\rho
\rho }$, $M_{xx}$, and $M_{\rho x}$ of the binary mixture. We will do the
calculation explicitly for $M_{\rho \rho }$ and quote the final result for
the other two elements. So from Eq. (\ref{intro:Mc1}) we find for $M_{\rho
\rho }$ 
\begin{eqnarray}
M_{\rho \rho } &=&x_{1}(1-\rho x_{1}\tilde{c}_{11})+x_{2}(1-\rho x_{2}\tilde{
c}_{22})-\rho x_{1}x_{2}(\tilde{c}_{12}+\tilde{c}_{21})  \notag \\
&=&V\rho \beta \sum_{i,j=1}^{2}x_{i}x_{j}\left( \frac{\partial \mu _{i}}{
\partial N_{j}}\right) _{T,V,\{N_{\bar{\jmath}}\}}  \notag \\
&=&\frac{\rho \beta }{\chi _{T}}\sum_{i,j=1}^{2}x_{i}x_{j}v_{i}v_{j}  \notag
\\
&=&\frac{\chi _{T}^{0}}{\chi _{T}}~,
\end{eqnarray}
where $\chi _{T}^{0}=\beta /\rho $ is the isothermal compressibility of the
ideal gas, in the second equality Eqs. (\ref{intro:tildeA}) and 
(\ref{app1:dmdn}) were used, in the third equality we used
Eqs. (\ref{app1:dmudN}) and (\ref{app1:G-D}) and in the last equality
Eq. (\ref{app1:V}). For $M_{\rho x}$ we find  
\begin{equation}
M_{\rho x}=\sqrt{x_{1}x_{2}}\;\delta \;\frac{\chi _{T}^{0}}{\chi _{T}}~,
\end{equation}
where 
\begin{equation}
\delta \equiv \rho (v_{1}-v_{2})=\frac{1}{V}\left( \frac{\partial V}{
\partial x_{1}}\right) _{T,P,N}~,
\end{equation}
and for $M_{xx}$ 
\begin{equation}
M_{xx}=x_{1}x_{2}\delta ^{2}\frac{\chi _{T}^{0}}{\chi _{T}}+x_{1}x_{2}\frac{
\chi _{T}^{0}}{V}\left( \frac{\partial ^{2}G}{\partial x_{1}^{2}}\right)
_{T,P,N}~.
\end{equation}
The determinant factorizes 
\begin{equation*}
\det (\mathbf{M})=\det (\mathbf{\tilde{A}})=[\det (\mathbf{\widehat{Q}}
)]^{2}=x_{1}x_{2}\frac{(\chi _{T}^{0})^{2}}{\chi _{T}V}\left( \frac{\partial
^{2}G}{\partial x_{1}^{2}}\right) _{T,P,N}~,
\end{equation*}
thus yielding Eq. (\ref{det-thermo}) in the main text.

\begin{acknowledgments}
This work was supported by the Italian MIUR (PRIN-COFIN 2004/2005).
\end{acknowledgments}
\bibliographystyle{apsrev}
\bibliography{paper_b}

\newcommand\minus{-}
\begin{thebibliography}{41}
\expandafter\ifx\csname natexlab\endcsname\relax\def\natexlab#1{#1}\fi
\expandafter\ifx\csname bibnamefont\endcsname\relax
  \def\bibnamefont#1{#1}\fi
\expandafter\ifx\csname bibfnamefont\endcsname\relax
  \def\bibfnamefont#1{#1}\fi
\expandafter\ifx\csname citenamefont\endcsname\relax
  \def\citenamefont#1{#1}\fi
\expandafter\ifx\csname url\endcsname\relax
  \def\url#1{\texttt{#1}}\fi
\expandafter\ifx\csname urlprefix\endcsname\relax\def\urlprefix{URL }\fi
\providecommand{\bibinfo}[2]{#2}
\providecommand{\eprint}[2][]{\url{#2}}

\bibitem[{\citenamefont{Rowlinson}(1969)}]{Rowlinson}
\bibinfo{author}{\bibfnamefont{J.~S.} \bibnamefont{Rowlinson}},
  \emph{\bibinfo{title}{Liquids and Liquid Mixtures}}
  (\bibinfo{publisher}{Butterworths}, \bibinfo{address}{London},
  \bibinfo{year}{1969}), \bibinfo{edition}{2nd} ed.

\bibitem[{\citenamefont{Lupis}(1983)}]{Lupis}
\bibinfo{author}{\bibfnamefont{C.~H.~P.} \bibnamefont{Lupis}},
  \emph{\bibinfo{title}{Chemical Thermodynamics of Materials}}
  (\bibinfo{publisher}{North-Holland}, \bibinfo{address}{Dordrecht},
  \bibinfo{year}{1983}).

\bibitem[{\citenamefont{Gazzillo}(1994)}]{Gazzillo94}
\bibinfo{author}{\bibfnamefont{D.}~\bibnamefont{Gazzillo}},
  \bibinfo{journal}{Mol. Phys.} \textbf{\bibinfo{volume}{{\bf 83}}},
  \bibinfo{pages}{1171} (\bibinfo{year}{1994}).

\bibitem[{\citenamefont{Gazzillo}(1995)}]{Gazzillo95}
\bibinfo{author}{\bibfnamefont{D.}~\bibnamefont{Gazzillo}},
  \bibinfo{journal}{Mol. Phys.} \textbf{\bibinfo{volume}{{\bf 84}}},
  \bibinfo{pages}{303} (\bibinfo{year}{1995}).

\bibitem[{\citenamefont{Chen and Forstmann}(1992)}]{Chen92}
\bibinfo{author}{\bibfnamefont{X.~S.} \bibnamefont{Chen}} \bibnamefont{and}
  \bibinfo{author}{\bibfnamefont{F.}~\bibnamefont{Forstmann}},
  \bibinfo{journal}{J. Chem. Phys.} \textbf{\bibinfo{volume}{{\bf 97}}},
  \bibinfo{pages}{3696} (\bibinfo{year}{1992}).

\bibitem[{\citenamefont{{C. P. Ursenbach and G. N. Patey}}(1994)}]{Ursenbach94}
\bibinfo{author}{\bibnamefont{{C. P. Ursenbach and G. N. Patey}}},
  \bibinfo{journal}{J. Chem. Phys.} \textbf{\bibinfo{volume}{{\bf 100}}},
  \bibinfo{pages}{3827} (\bibinfo{year}{1994}).

\bibitem[{\citenamefont{Bhatia and Thornton}(1970)}]{Bhatia70}
\bibinfo{author}{\bibfnamefont{A.~B.} \bibnamefont{Bhatia}} \bibnamefont{and}
  \bibinfo{author}{\bibfnamefont{D.~E.} \bibnamefont{Thornton}},
  \bibinfo{journal}{Phys. Rev. B} \textbf{\bibinfo{volume}{{\bf 2}}},
  \bibinfo{pages}{3004} (\bibinfo{year}{1970}).

\bibitem[{\citenamefont{Baxter}(1968)}]{Baxter68}
\bibinfo{author}{\bibfnamefont{R.~J.} \bibnamefont{Baxter}},
  \bibinfo{journal}{J. Chem. Phys.} \textbf{\bibinfo{volume}{{\bf 49}}},
  \bibinfo{pages}{2770} (\bibinfo{year}{1968}).

\bibitem[{\citenamefont{Baxter}()}]{Baxter71}
\bibinfo{author}{\bibfnamefont{R.~J.} \bibnamefont{Baxter}}, \bibinfo{note}{{in
  {\sl Physical Chemistry, an Advanced Treatise}, vol. 8A, ed. D. Henderson
  (Academic Press, New York, 1971) ch. 4}}.

\bibitem[{\citenamefont{{R. O. Watts, D. Henderson, and R. J.
  Baxter}}(1971)}]{Watts71}
\bibinfo{author}{\bibnamefont{{R. O. Watts, D. Henderson, and R. J. Baxter}}},
  \bibinfo{journal}{Advan. Chem. Phys.} \textbf{\bibinfo{volume}{{\bf 21}}},
  \bibinfo{pages}{421} (\bibinfo{year}{1971}).

\bibitem[{\citenamefont{Baxter}(1970)}]{Baxter70}
\bibinfo{author}{\bibfnamefont{R.~J.} \bibnamefont{Baxter}},
  \bibinfo{journal}{J. Chem. Phys.} \textbf{\bibinfo{volume}{{\bf 52}}},
  \bibinfo{pages}{4559} (\bibinfo{year}{1970}).

\bibitem[{\citenamefont{Barboy}(1975)}]{Barboy74}
\bibinfo{author}{\bibfnamefont{B.}~\bibnamefont{Barboy}},
  \bibinfo{journal}{Chem. Phys.} \textbf{\bibinfo{volume}{{\bf 11}}},
  \bibinfo{pages}{357} (\bibinfo{year}{1975}).

\bibitem[{\citenamefont{Perram and Smith}(1975)}]{Perram75}
\bibinfo{author}{\bibfnamefont{J.~W.} \bibnamefont{Perram}} \bibnamefont{and}
  \bibinfo{author}{\bibfnamefont{E.~R.} \bibnamefont{Smith}},
  \bibinfo{journal}{Chem. Phys. Lett.} \textbf{\bibinfo{volume}{{\bf 35}}},
  \bibinfo{pages}{138} (\bibinfo{year}{1975}).

\bibitem[{\citenamefont{Barboy and Tenne}(1979)}]{Barboy79}
\bibinfo{author}{\bibfnamefont{B.}~\bibnamefont{Barboy}} \bibnamefont{and}
  \bibinfo{author}{\bibfnamefont{R.}~\bibnamefont{Tenne}},
  \bibinfo{journal}{Chem. Phys.} \textbf{\bibinfo{volume}{{\bf 38}}},
  \bibinfo{pages}{369} (\bibinfo{year}{1979}).

\bibitem[{\citenamefont{Stell}(1991)}]{Stell91}
\bibinfo{author}{\bibfnamefont{G.}~\bibnamefont{Stell}}, \bibinfo{journal}{J.
  Stat. Phys.} \textbf{\bibinfo{volume}{{\bf 63}}}, \bibinfo{pages}{1203}
  (\bibinfo{year}{1991}).

\bibitem[{\citenamefont{{C. Robertus, W. H. Philipse, J. G. H. Joosten, and Y.
  K. Levine}}(1989)}]{Robertus89}
\bibinfo{author}{\bibnamefont{{C. Robertus, W. H. Philipse, J. G. H. Joosten,
  and Y. K. Levine}}}, \bibinfo{journal}{J. Chem. Phys.}
  \textbf{\bibinfo{volume}{{\bf 90}}}, \bibinfo{pages}{4482}
  (\bibinfo{year}{1989}).

\bibitem[{\citenamefont{{S. H. Chen, J. Rouch, F. Sciortino, and P.
  Tartaglia}}(1994)}]{Chen94}
\bibinfo{author}{\bibnamefont{{S. H. Chen, J. Rouch, F. Sciortino, and P.
  Tartaglia}}}, \bibinfo{journal}{J. Phys.:Condens. Matter}
  \textbf{\bibinfo{volume}{{\bf 6}}}, \bibinfo{pages}{10855}
  (\bibinfo{year}{1994}).

\bibitem[{\citenamefont{{H. L\"owen}}(1994)}]{Lowen94}
\bibinfo{author}{\bibnamefont{{H. L\"owen}}}, \bibinfo{journal}{Phys. Rep.}
  \textbf{\bibinfo{volume}{{\bf 237}}}, \bibinfo{pages}{249}
  (\bibinfo{year}{1994}).

\bibitem[{\citenamefont{N\"agele}(1996)}]{Nagele96}
\bibinfo{author}{\bibfnamefont{G.}~\bibnamefont{N\"agele}},
  \bibinfo{journal}{Phys. Rep.} \textbf{\bibinfo{volume}{{\bf 272}}},
  \bibinfo{pages}{215} (\bibinfo{year}{1996}).

\bibitem[{\citenamefont{Gazzillo and Giacometti}(2000)}]{Gazzillo00}
\bibinfo{author}{\bibfnamefont{D.}~\bibnamefont{Gazzillo}} \bibnamefont{and}
  \bibinfo{author}{\bibfnamefont{A.}~\bibnamefont{Giacometti}},
  \bibinfo{journal}{J. Chem. Phys.} \textbf{\bibinfo{volume}{{\bf 113}}},
  \bibinfo{pages}{9837} (\bibinfo{year}{2000}).

\bibitem[{\citenamefont{Gazzillo and Giacometti}(2004)}]{Gazzillo04}
\bibinfo{author}{\bibfnamefont{D.}~\bibnamefont{Gazzillo}} \bibnamefont{and}
  \bibinfo{author}{\bibfnamefont{A.}~\bibnamefont{Giacometti}},
  \bibinfo{journal}{J. Chem. Phys.} \textbf{\bibinfo{volume}{{\bf 120}}},
  \bibinfo{pages}{4742} (\bibinfo{year}{2004}).

\bibitem[{fn1()}]{fn1}
\bibinfo{note}{Note that the mMSA solution for the Baxter factor function of
  the SHS binary model can also be obtained \cite{Gazzillo03} from the model of
  a mixture of hard spheres interacting through Yukawa potentials in the MSA
  \cite{Blum78} upon taking the limit of a vanishing screening length.}

\bibitem[{\citenamefont{{M. H. G. M. Penders and A. Vrij}}(1991)}]{Penders91}
\bibinfo{author}{\bibnamefont{{M. H. G. M. Penders and A. Vrij}}},
  \bibinfo{journal}{Physica A} \textbf{\bibinfo{volume}{{\bf 173}}},
  \bibinfo{pages}{532} (\bibinfo{year}{1991}).

\bibitem[{\citenamefont{{C. Regnaut, S. Amokrane, and Y.
  Heno}}(1995)}]{Regnaut95}
\bibinfo{author}{\bibnamefont{{C. Regnaut, S. Amokrane, and Y. Heno}}},
  \bibinfo{journal}{J. Chem. Phys.} \textbf{\bibinfo{volume}{{\bf 102}}},
  \bibinfo{pages}{6230} (\bibinfo{year}{1995}).

\bibitem[{\citenamefont{{S. Amokrane and C. Regnaut}}(1997)}]{Amokrane97}
\bibinfo{author}{\bibnamefont{{S. Amokrane and C. Regnaut}}},
  \bibinfo{journal}{J. Chem. Phys.} \textbf{\bibinfo{volume}{{\bf 106}}},
  \bibinfo{pages}{376} (\bibinfo{year}{1997}).

\bibitem[{\citenamefont{{R. Fantoni, D. Gazzillo, and A.
  Giacometti}}(2005)}]{Fantoni05}
\bibinfo{author}{\bibnamefont{{R. Fantoni, D. Gazzillo, and A. Giacometti}}},
  \bibinfo{journal}{J. Chem. Phys.} \textbf{\bibinfo{volume}{{\bf 122}}},
  \bibinfo{pages}{034901} (\bibinfo{year}{2005}).

\bibitem[{\citenamefont{Caillol}(2002)}]{Caillol02}
\bibinfo{author}{\bibfnamefont{J.~M.} \bibnamefont{Caillol}},
  \bibinfo{journal}{J. Phys. A-Math. Gen.} \textbf{\bibinfo{volume}{{\bf 35}}},
  \bibinfo{pages}{4189} (\bibinfo{year}{2002}).

\bibitem[{\citenamefont{{L. D. Landau and E. M. Lifshitz}}(2001)}]{Landau}
\bibinfo{author}{\bibnamefont{{L. D. Landau and E. M. Lifshitz}}},
  \emph{\bibinfo{title}{Statistical Physics}}
  (\bibinfo{publisher}{Butterworth-Heinemann}, \bibinfo{year}{2001}),
  \bibinfo{edition}{3rd} ed., \bibinfo{note}{part 1}.

\bibitem[{fn2()}]{fn2}
\bibinfo{note}{Note that in the mMSA, unlike in the PY, $y_{ij}(\sigma
  _{ij}^{+})$ differs from $y_{ij}(R_{ij}^{+})$ even in the sticky limit, in
  view of the fact that $y_{ij}(r)$ is not continuous at $R_{ij}$. This feature
  allows to obtain the correct HS result in the limit of no adhesion.}

\bibitem[{fn3()}]{fn3}
\bibinfo{note}{Our Eq. (\ref{detQSHS}) corrects Eq. (60) of Barboy and Tenne
  \cite{Barboy79} where a sign is misprinted. Note also that these authors use
  the symbol $d_{ij}$ in place of our $\sigma_{ij}$ and
  $\tau_{ij}^{BT}=6\tau_{ij}$ where the superscript $BT$ indicates that the
  symbol is the one used by Barboy and Tenne.}

\bibitem[{\citenamefont{{N. B. Wilding, F. Schmid, and P.
  Nielaba}}(1998)}]{Wilding98}
\bibinfo{author}{\bibnamefont{{N. B. Wilding, F. Schmid, and P. Nielaba}}},
  \bibinfo{journal}{Phys. Rev. E} \textbf{\bibinfo{volume}{{\bf 58}}},
  \bibinfo{pages}{2201} (\bibinfo{year}{1998}).

\bibitem[{\citenamefont{{S. Asakura and F. Oosawa}}(1954)}]{Asakura54}
\bibinfo{author}{\bibnamefont{{S. Asakura and F. Oosawa}}},
  \bibinfo{journal}{J. Chem. Phys.} \textbf{\bibinfo{volume}{{\bf 22}}},
  \bibinfo{pages}{1255} (\bibinfo{year}{1954}).

\bibitem[{\citenamefont{{J. L. Barrat and J. P. Hansen}}(2003)}]{Hansen-Barrat}
\bibinfo{author}{\bibnamefont{{J. L. Barrat and J. P. Hansen}}},
  \emph{\bibinfo{title}{Basic Concepts for Simple and Complex Liquids}}
  (\bibinfo{publisher}{Cambridge}, \bibinfo{year}{2003}).

\bibitem[{\citenamefont{Sollich}(2005)}]{Sollich05}
\bibinfo{author}{\bibfnamefont{P.}~\bibnamefont{Sollich}}
  (\bibinfo{year}{2005}), \bibinfo{note}{after the submission of this work it
  has been brought to our attention [P. Sollich, private communication] that
  Cases I, III, and V have a pair interaction potential [see Eq. (\ref{SW})]
  which might include many-body effects due to the presence of the averages.
  This is bacause, for instance, the interaction between two particles of
  species 1 depends also on the diameter of the particles of species 2. As a
  complement to this observation, we also note that as the diameter of one
  species of particles, say $\sigma_1\to 0$, vanishes, then -according to Eq.
  (\ref{SW})- those pointwise particles are non-interacting among themselves
  thus requiring $K_{11}\to 0$ as well. This constraint would rule out Cases I
  and III.}

\bibitem[{\citenamefont{{J. L. Lebowitz and J. S.
  Rowlinson}}(1964)}]{Lebowitz64}
\bibinfo{author}{\bibnamefont{{J. L. Lebowitz and J. S. Rowlinson}}},
  \bibinfo{journal}{J. Chem. Phys.} \textbf{\bibinfo{volume}{{\bf 41}}},
  \bibinfo{pages}{133} (\bibinfo{year}{1964}).

\bibitem[{\citenamefont{{T. Biben and J.-P. Hansen}}(1991)}]{Biben91}
\bibinfo{author}{\bibnamefont{{T. Biben and J.-P. Hansen}}},
  \bibinfo{journal}{Phys. Rev. Lett.} \textbf{\bibinfo{volume}{{\bf 66}}},
  \bibinfo{pages}{2215} (\bibinfo{year}{1991}).

\bibitem[{\citenamefont{{C. J\c{e}drzejek, J. Konior, and M.
  Streszewski}}(1987)}]{Jedrzejek87}
\bibinfo{author}{\bibnamefont{{C. J\c{e}drzejek, J. Konior, and M.
  Streszewski}}}, \bibinfo{journal}{Phys. Rev. A} \textbf{\bibinfo{volume}{{\bf
  35}}}, \bibinfo{pages}{1226} (\bibinfo{year}{1987}).

\bibitem[{\citenamefont{{O. Antonevych, F. Forstmann, and E.
  Diaz-Herrera}}(2002)}]{Antonevych02}
\bibinfo{author}{\bibnamefont{{O. Antonevych, F. Forstmann, and E.
  Diaz-Herrera}}}, \bibinfo{journal}{Phys. Rev. E}
  \textbf{\bibinfo{volume}{{\bf 65}}}, \bibinfo{pages}{061504}
  (\bibinfo{year}{2002}).

\bibitem[{\citenamefont{Ashcroft and Langreth}(1967)}]{Ashcroft67}
\bibinfo{author}{\bibfnamefont{N.~W.} \bibnamefont{Ashcroft}} \bibnamefont{and}
  \bibinfo{author}{\bibfnamefont{D.~C.} \bibnamefont{Langreth}},
  \bibinfo{journal}{Phys. Rev.} \textbf{\bibinfo{volume}{{\bf 156}}},
  \bibinfo{pages}{685} (\bibinfo{year}{1967}).

\bibitem[{\citenamefont{Gazzillo and Giacometti}(2003)}]{Gazzillo03}
\bibinfo{author}{\bibfnamefont{D.}~\bibnamefont{Gazzillo}} \bibnamefont{and}
  \bibinfo{author}{\bibfnamefont{A.}~\bibnamefont{Giacometti}},
  \bibinfo{journal}{Mol. Phys.} \textbf{\bibinfo{volume}{{\bf 101}}},
  \bibinfo{pages}{2171} (\bibinfo{year}{2003}).

\bibitem[{\citenamefont{{L. Blum and J. S. H{\o}ye}}(1978)}]{Blum78}
\bibinfo{author}{\bibnamefont{{L. Blum and J. S. H{\o}ye}}},
  \bibinfo{journal}{J. Stat. Phys.} \textbf{\bibinfo{volume}{{\bf 19}}},
  \bibinfo{pages}{317} (\bibinfo{year}{1978}).

\end{thebibliography}

\newpage \centerline{\bf LIST OF FIGURES}

\begin{figure}[h!]
\begin{center}
\includegraphics[width=10cm]{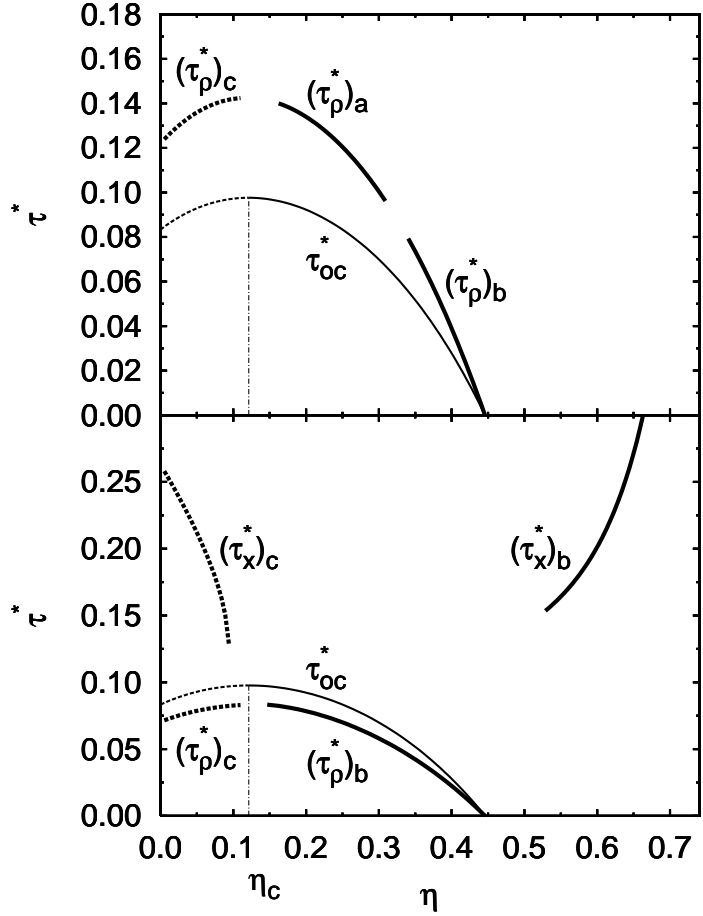}
\end{center}
\caption{Spinodal line for the symmetric mixture in the PY approximation
with $\protect\nu_{12}=2$ in the top panel and with
$\protect\nu_{12}=2/3$ in the bottom panel. At $\protect\nu_{12}=2$ the
instability is of pure condensation 
along $\protect\tau^*=(\protect\tau^*_\protect\protect\rho)_i$,
$i=a,b$ and of pure demixing along
$\protect\tau^*=(\protect\tau^*_x)_i$, $i=a,b$. The zeroes 
labeled $c$ are unphysical. The gaps between the curves $\protect\tau^*=(
\protect\tau^*_\protect\protect\rho)_i$ are numerical artifacts. At
$\protect\nu_{12}=2/3$ there is the appearance of a pure demixing
spinodal at high $\protect\eta$ which does not cross the pure
condensation one. 
For reference we also plot in both panels the spinodal of the one
component system $\protect\tau^*_{oc}$ [see
Eq. (\protect\ref{py-oc-spinodal})] which is physical only for  
$\protect\eta>\protect\eta_c=(3\protect\sqrt{2}-4)/2=0.1213\ldots$.}
\label{fig:py-sym1}
\end{figure}

\begin{figure}[h!]
\begin{center}
\includegraphics[width=10cm]{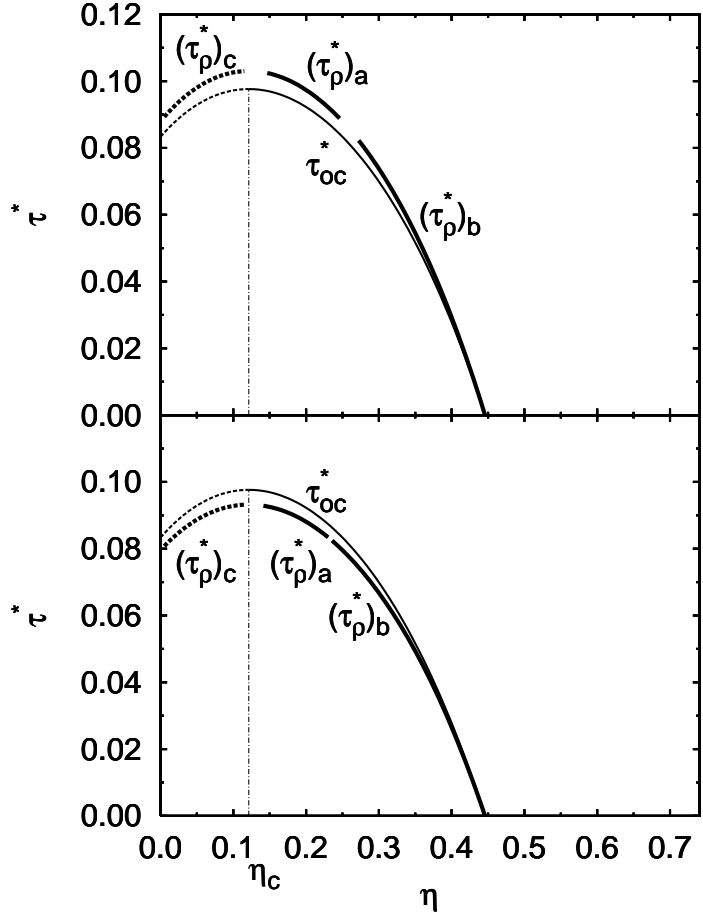}
\end{center}
\caption{Same as Fig. \protect\ref{fig:py-sym1} with
$\protect\nu_{12}=1/0.9$ in the top panel and
$\protect\nu_{12}=1/1.1$ in the bottom panel. In this last case the
expected line of pure demixing would start at $\protect 
\eta>\protect\eta_0=0.7404\ldots$ in the unphysical range of
densities.}  
\label{fig:py-sym1-oclim}
\end{figure}

\begin{figure}[h!]
\begin{center}
\includegraphics[width=10cm]{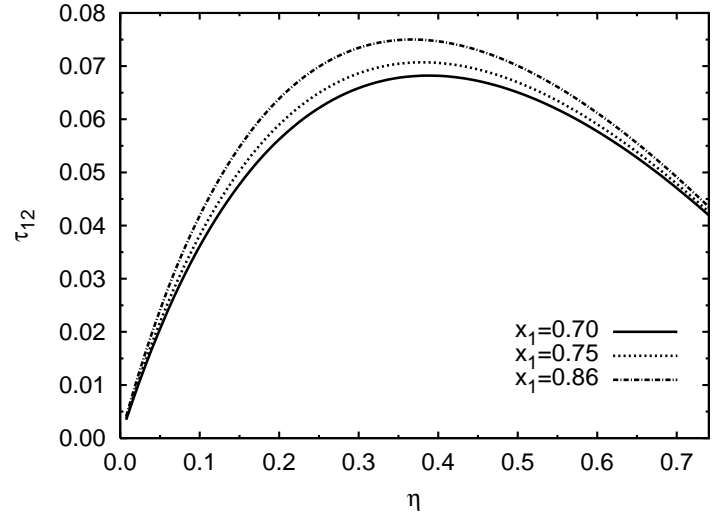}
\end{center}
\caption{For System A the mMSA spinodal [see
Eq. (\protect\ref{A:spinodal-mMSA})] for $\protect\zeta =2$ and three
different values of $x_{1}$. } 
\label{fig:system-a-mmsa-t}
\end{figure}

\begin{figure}[h!]
\begin{center}
\includegraphics[width=10cm]{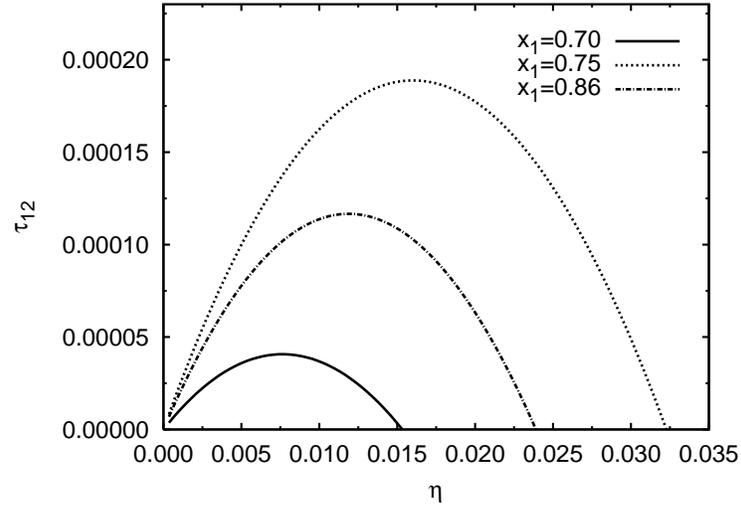}
\end{center}
\caption{For System A the PY spinodal [see Eq. (\protect\ref{A:spinodal-PY}
)] under the same conditions considered in
Fig. \protect\ref{fig:system-a-mmsa-t}.} 
\label{fig:system-a-py-t}
\end{figure}

\begin{figure}[h!]
\begin{center}
\includegraphics[width=10cm]{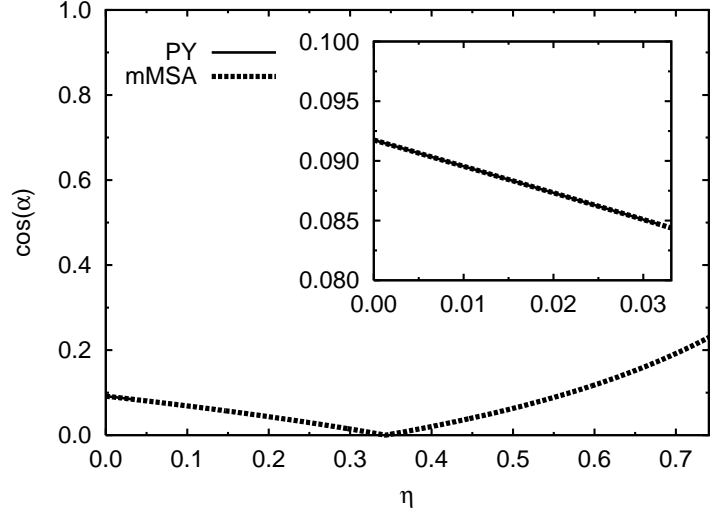}
\end{center}
\caption{Behavior of the angle $\protect\alpha $ of
Eq. (\protect\ref{intro:angle}) predicted by the mMSA and PY for
System A when $x_{1}=0.75$, $\protect\zeta =2$. In this case the PY
spinodal has no solutions when $\protect\eta >0.03227\ldots $. In the
inset we show the region of $\protect\eta $ were the PY spinodal
exists. Note that here and in the following $\cos \protect\alpha $
rather than the angle $\protect\alpha $ itself is depicted for visual
convenience.} 
\label{fig:system-a-a}
\end{figure}

\begin{figure}[h!]
\begin{center}
\includegraphics[width=10cm]{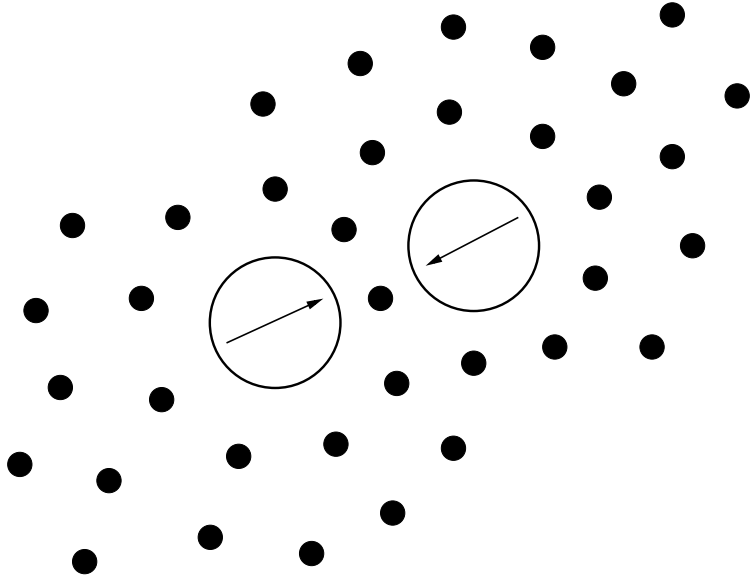}
\end{center}
\caption{In System B, when we have a small number of large particles of
species 1, the demixing instability [see Eq. (\protect\ref{B-od})]
should be favored by the osmotic depletion mechanism, since the small
spheres interact through a HS potential both among themselves and with
the big spheres.} 
\label{fig:od}
\end{figure}

\begin{figure}[h!]
\begin{center}
\includegraphics[width=10cm]{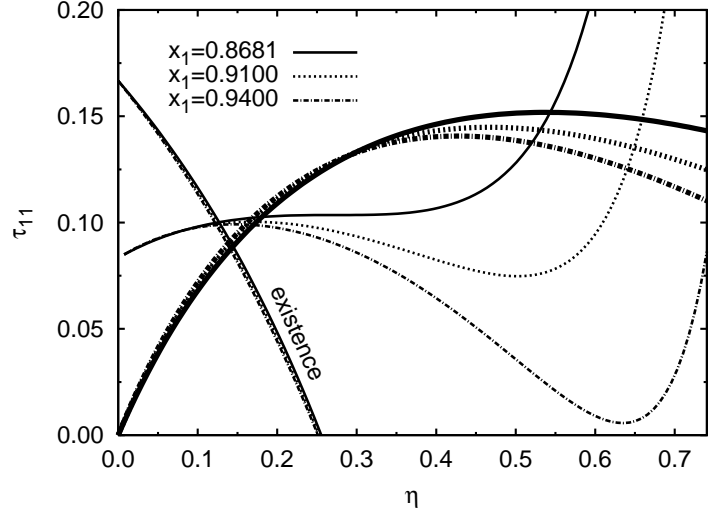}
\end{center}
\caption{For System B the spinodals predicted by mMSA [thick lines,
see Eq. (\protect\ref{B:spinodal-mMSA})] and the ones predicted by PY
[thin lines, see Eq. (\protect\ref{B:spinodal-PY})] for $\protect\zeta
=1$ at three 
different values of $x_{1}$. The physically meaningful PY spinodals are
those lying above the \textquotedblleft existence\textquotedblright\ lines
in accord with condition (\protect\ref{sysB:PY-sp-ex}). }
\label{fig:system-b-py-t}
\end{figure}

\begin{figure}[h!]
\begin{center}
\includegraphics[width=10cm]{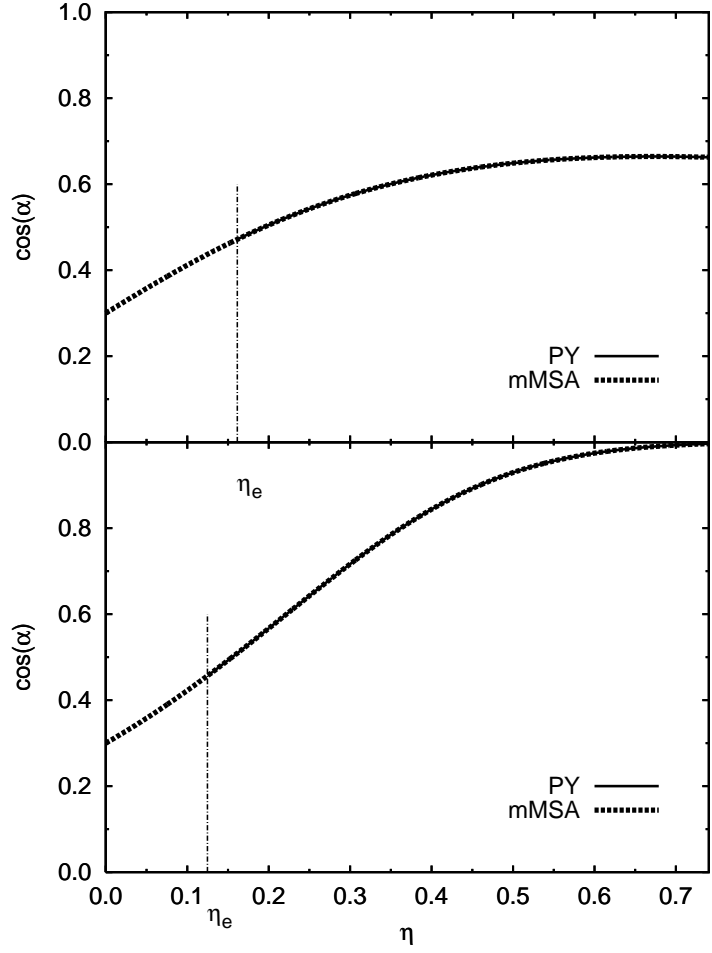}
\end{center}
\caption{For System B behavior of the angle $\protect\alpha$ of Eq. 
(\protect\ref{intro:angle}) predicted by mMSA and PY for $x_{1}=0.91$
and $\protect\zeta=1$ in the bottom panel (in this case the PY
spinodal has loss of solution for $\protect\eta <\protect\eta
_{e}\approx 0.1248\ldots $.) and $\protect\zeta=2$ in the top panel
(in this case the PY spinodal has loss of solution for $\protect
\eta <\protect\eta _{e}\approx 0.1614\ldots $).}
\label{fig:system-b-py-a-y1}
\end{figure}

\begin{figure}[h!]
\begin{center}
\includegraphics[width=10cm]{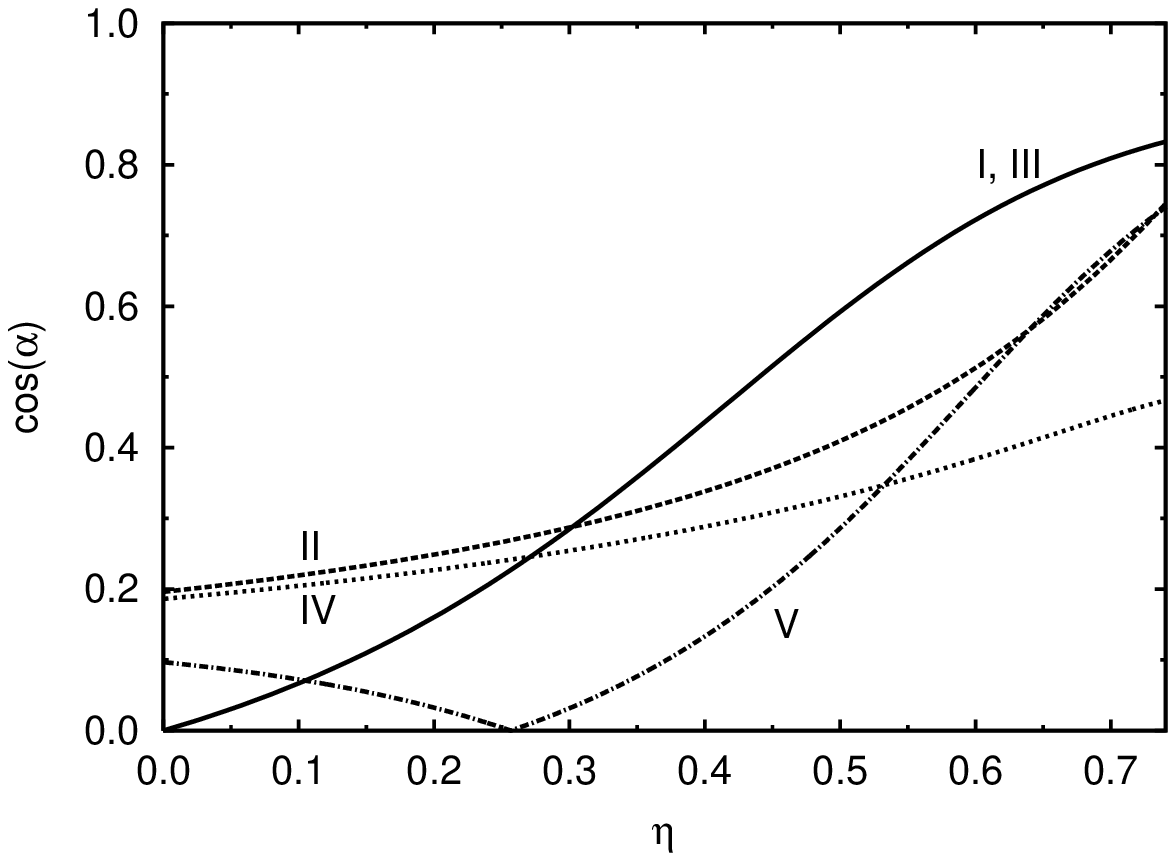}
\end{center}
\caption{Behavior of the angle $\protect\alpha$ of
Eq. (\protect\ref{intro:angle}) for Cases I, II, III, IV, and V when
$x_1=1/2$ and $\protect\zeta=3/2$.} 
\label{fig:Cases1}
\end{figure}

\begin{figure}[h!]
\begin{center}
\includegraphics[width=10cm]{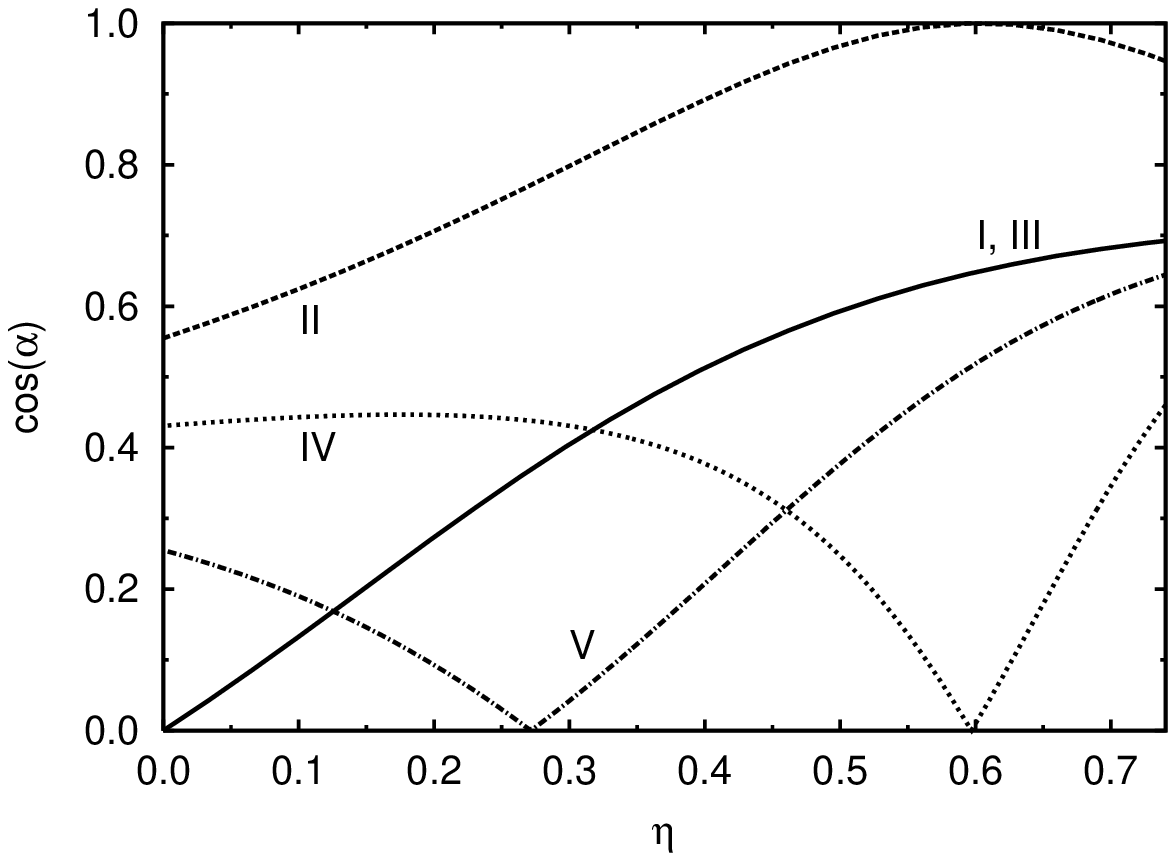}
\end{center}
\caption{Behavior of the angle $\protect\alpha$ of
Eq. (\protect\ref{intro:angle}) for Cases I, II, III, IV, and V when
$x_1=1/2$ and $\protect\zeta=5$.} 
\label{fig:Cases2}
\end{figure}

\begin{itemize}
\item[Fig. 1] Schematic representation of the two orthonormal vectors
${\bf z}_\pm$ defined in Eq. (\ref{zpm}) and of the angle 
$\alpha$ defined in Eq. (\ref{intro:angle}) when
$\alpha\in[0,\pi/2]$. When $\alpha\in[-\pi/2,0]$ the angle
shown in the Fig. corresponds to $|\alpha|$ and $\delta\bar{x}$ to
$-\delta\bar{x}$.

\item[Fig. 2] Spinodal line (continuous curve) for the symmetric mixture in the
mMSA with $\protect\nu _{12}=2$. The kind of instability is of pure
condensation along the whole spinodal.

\item[Fig. 3] Spinodal line (continuous curve) for the symmetric mixture in the
mMSA with $\protect\nu _{12}=2/3$. In this case the instability is of pure
condensation for $\protect\eta <\protect\eta _{\protect\rho x}$ along 
$\protect\tau ^{\ast }=\protect\tau _{\rho}^{\ast }$ and of pure
demixing for $\protect\eta >\protect\eta _{\protect\rho x}$ along $\protect
\tau ^{\ast }=\protect\tau _{x}^{\ast }$.

\item[Fig. 4] Spinodal line for the symmetric mixture in the PY approximation
with $\protect\nu_{12}=2$ in the top panel and with
$\protect\nu_{12}=2/3$ in the bottom panel. At $\protect\nu_{12}=2$ the
instability is of pure condensation 
along $\protect\tau^*=(\protect\tau^*_\protect\protect\rho)_i$,
$i=a,b$ and of pure demixing along
$\protect\tau^*=(\protect\tau^*_x)_i$, $i=a,b$. The zeroes 
labeled $c$ are unphysical. The gaps between the curves $\protect\tau^*=(
\protect\tau^*_\protect\protect\rho)_i$ are numerical artifacts. At
$\protect\nu_{12}=2/3$ there is the appearance of a pure demixing
spinodal at high $\protect\eta$ which does not cross the pure
condensation one. 
For reference we also plot in both panels the spinodal of the one
component system $\protect\tau^*_{oc}$ [see
Eq. (\protect\ref{py-oc-spinodal})] which is physical only for  
$\protect\eta>\protect\eta_c=(3\protect\sqrt{2}-4)/2=0.1213\ldots$.

\item[Fig. 5] Same as Fig. \protect\ref{fig:py-sym1} with
$\protect\nu_{12}=1/0.9$ in the top panel and
$\protect\nu_{12}=1/1.1$ in the bottom panel. In this last case the
expected line of pure demixing would start at $\protect 
\eta>\protect\eta_0=0.7404\ldots$ in the unphysical range of
densities.

\item[Fig. 6] For System A the mMSA spinodal [see
Eq. (\protect\ref{A:spinodal-mMSA})] for $\protect\zeta =2$ and three
different values of $x_{1}$. 

\item[Fig. 7] For System A the PY spinodal [see
Eq. (\protect\ref{A:spinodal-PY})] under the same conditions
considered in Fig. \protect\ref{fig:system-a-mmsa-t}. 

\item[Fig. 8] Behavior of the angle $\protect\alpha $ of
Eq. (\protect\ref{intro:angle}) predicted by the mMSA and PY for
System A when $x_{1}=0.75$, $\protect\zeta =2$. In this case the PY
spinodal has no solutions when $\protect\eta >0.03227\ldots $. In the
inset we show the region of $\protect\eta $ were the PY spinodal
exists. Note that here and in the following $\cos \protect\alpha $
rather than the angle $\protect\alpha $ itself is depicted for visual
convenience.  

\item[Fig. 9] In System B, when we have a small number of large
particles of species 1, the demixing instability [see
Eq. (\protect\ref{B-od})] should be favored by the osmotic depletion
mechanism, since the small spheres interact through a HS potential
both among themselves and with the big spheres.

\item[Fig. 10] For System B the spinodals predicted by mMSA [thick
lines, see Eq. (\protect\ref{B:spinodal-mMSA})] and the ones
predicted by PY [thin lines, 
see Eq. (\protect\ref{B:spinodal-PY})] for $\protect\zeta =1$ at three
different values of $x_{1}$. The physically meaningful PY spinodals are
those lying above the \textquotedblleft existence\textquotedblright\ lines
in accord with condition (\protect\ref{sysB:PY-sp-ex}).

\item[Fig. 11] For System B behavior of the angle $\protect\alpha$ of Eq. 
(\protect\ref{intro:angle}) predicted by mMSA and PY for $x_{1}=0.91$
and $\protect\zeta=1$ in the bottom panel (in this case the PY
spinodal has loss of solution for $\protect\eta <\protect\eta
_{e}\approx 0.1248\ldots $.) and $\protect\zeta=2$ in the top panel
(in this case the PY spinodal has loss of solution for $\protect
\eta <\protect\eta _{e}\approx 0.1614\ldots $).

\item[Fig. 12] Behavior of the angle $\protect\alpha$ of
Eq. (\protect\ref{intro:angle}) for Cases I, II, III, IV, and V when
$x_1=1/2$ and $\protect\zeta=3/2$. 

\item[Fig. 13] Behavior of the angle $\protect\alpha$ of Eq. 
(\protect\ref{intro:angle}) for Cases I, II, III, IV, and V when
$x_1=1/2$ and $\protect\zeta=5$. 

\end{itemize}

\newpage

\end{document}